\documentclass[a4paper,11pt]{article}
\pdfoutput=1 

\usepackage{jheppub} 

\usepackage[T1]{fontenc} 
\usepackage{mathrsfs}

\def\L{\Lambda}

\def\k{\kappa}
\def\d{\delta}

\def\l{\lambda}

\def\v{\varrho}

\def\nprod{\prod_{n\neq 0}}
\def\nplus{\prod_{n>0}}
\def\nkprod{\prod_{n\neq 0,k}}
\def\dT{\delta T}
\def\cl{^{cl}}
\def\nprod{\prod_{n\neq 0}}
\def\nplus{\prod_{n>0}}
\def\nkprod{\prod_{n\neq 0,k}}
\def\n{\mathcal{N}}
\def\L{\mathcal L}
\def\Lt{\tilde{\mathcal{L}}}
\def\v{\mathfrak{v}}
\def\z{\mathfrak{z}}
\def\y{\mathfrak{y}}
\def\det{\text{det}}
\def\pe{\pi \epsilon}
\def\tpa{2\pi\alpha'}
\def\sech{\mathrm{sech}}
\def\k{^{(k)}}
\def\E{\mathscr{E}}
\newcommand{\deriv}[2]{\frac{\mathrm{d}#1}{\mathrm{d}#2}}
\newcommand{\modd}[1]{\|#1\|}
\newcommand{\ev}[1]{\langle #1 \rangle}

\title{\boldmath{Worldsheet Instantons and the amplitude for string pair 
production in an external field as a WKB exact functional  integral}}

\author { James Gordon and  Gordon W. Semenoff}

\affiliation {Department of Physics and Astronomy, University of
British Columbia,  6224 Agricultural Road, Vancouver, British Columbia, Canada V6T 1Z1}

\emailAdd{jbgordon@phas.ubc.ca}
\emailAdd{gordonws@phas.ubc.ca}

\abstract{We revisit the problem of charged string pair creation in a constant external electric field. The string states are massive and creation of pairs from the vacuum is a tunnelling process, analogous to the Schwinger process where charged particle-anti-particle pairs are created by an electric field.  
We find the instantons in the worldsheet sigma model which are responsible for the tunnelling events.  We evaluate the sigma model partition function in the multi-instanton sector in the WKB approximation which keeps the classical action and integrates the quadratic fluctuations about the solution.   We find that the summation 
of the result over all multi-instanton sectors reproduces the known amplitude.  This suggests that corrections to the WKB limit must cancel.  To show
that they indeed cancel, we identify a fermionic symmetry of the sigma model which occurs in the instanton sectors and which is associated with collective coordinates.   
We demonstrate that the action is symmetric and  that the interaction action is an exact form.  These conditions are sufficient for localization of the worldsheet functional integral onto its WKB limit. }

\begin{document} 
\maketitle
\flushbottom

\section{Introduction}
Schwinger's famous formula \cite{Schwinger1951} for the rate of vacuum decay, per unit volume per unit time, by the production of charged particle-anti-particle pairs in a constant electric field is given by 
\begin{align}\label{particleschwinger}
\Gamma_{\rm particle} = (2J+1)\pi \sum_{k=1}^\infty (-1)^{(k+1)(2J+1)} \left[\frac{E}{4\pi^2  ~k}\right]^{\frac{D+1}{2}}e^{-\frac{\pi m^2}{E}k}
\end{align}
where $E$ is the strength of the electric field and $J$ and $m$ are the spin and mass of the particle.   We have
generalized the usual formula to $D$ space dimensions and we have absorbed the electric charge into the field $E$.   
This is one of the few non-perturbative formulae describing a quantum field theory process which 
can be obtained exactly, normally by computing the imaginary part of the vacuum energy, and thereby the vacuum decay rate, 
when a constant electric field is present.  

There exists a formula analogous to equation (\ref{particleschwinger})  for the pair production of electrically charged strings by a constant electric field.    
In that case, the charges reside on the endpoints of open strings which are in turn confined to travel on D-branes. 
When the internal electric field on an infinite, flat D-brane has strength $E$, and assuming that the other end of the string goes to a parallel D-brane with no electric field, the formula for the rate of vacuum decay due to string pair creation, per unit volume per unit time is \cite{Bachas1992,Ambjorn2003}
\begin{align}\label{stringschwinger}
\Gamma_{\rm string} =\sum_S~\frac{  E }{\mathcal E} \pi  \sum_{k=1}^\infty
(-1)^{k+1}\left[  \frac{\mathcal E }{4\pi^2 k}\right]^{\frac{D+1}{2}} e^{-\frac{\pi m_S^2}
{\mathcal E}k-2 \pi  \alpha'{\mathcal Ek}}
\end{align}
where
\begin{align}\label{epsilon}
{\mathcal E}=  \frac{{\rm arctanh} 2\pi\alpha' E}{2\pi\alpha'}\sim E\left[1+{\mathcal O} (2\pi\alpha'E)^2 \right]
\end{align}
The sum over $S$ is the sum over the particle states in the string spectrum.  
The $m_S$ are the masses of the particles and the multiplicity of the states 
with each mass is analogous to the $(2J+1)$ factor in the Schwinger formula (\ref{particleschwinger}).

The string formula (\ref{stringschwinger}) reduces to  the particle formula (\ref{particleschwinger}), summed over the particles in the 
string spectrum,  in the limit
where the electric field is much smaller than the string tension, $2\pi\alpha' E<<1$ so that ${\mathcal E}\approx E$.  Away from that limit, 
the string formula (\ref{stringschwinger}) is not simply identical to the particle formula  (\ref{particleschwinger}) summed
over the particle states in the string spectrum.  
The main (but not only) difference is the replacement of the electric field $E$ by the parameter ${\mathcal E}$,
which can be thought of as a type of screening.  
Moreover, since ${\rm arctanh}2\pi \alpha' E $ becomes complex if $2\pi\alpha' E>1$, there is an upper 
critical electric field,
\begin{align}\label{critical_electric_field}
E_{\rm crit.} ~\equiv~\frac{1}{2\pi\alpha'}
\end{align}
which agrees with other derivations of the upper critical electric field, being simply the 
value of the electric field which balances the string tension on flat space. It can already be
seen to be the singular point in the Born-Infeld action\footnote{
The leading terms in an expansion in derivatives of $F_{\mu\nu}$ of the disc amplitude 
is the Born-Infeld effective action
$$
S_{DBI}= \frac{1}{(2\pi)^{D}{\alpha'}^{(D+1)/2} g_s}\int d^{D+1}\xi \sqrt{\det\left(g_{\mu\nu}(\xi)+2\pi\alpha' F_{\mu\nu}(\xi)\right)}
$$ integrated over the D-brane world-volume.   For a flat brane, $g_{\mu\nu}=\delta_{\mu\nu}$ and constant electric field, where 
the non-zero components
of $F_{\mu\nu}$ are 
$F_{01}=-F_{10}=E$, the Born-Infeld action does not have an imaginary part, as long as the electric field is less than the critical field $E\leq\frac{1}{2\pi\alpha'}$. 
In string perturbation theory about flat space, and with $E\leq\frac{1}{2\pi\alpha'}$, the imaginary part of the vacuum energy first appears in the cylinder amplitude.  As we see in equations (\ref{stringschwinger}),(\ref{critical_electric_field}), the upper critical electric field also appears in the cylinder amplitude.}
which is contained in the disc
amplitude for the string sigma model \cite{Fradkin1985,Fradkin1985a,Fradkin1985b}.  
Equation (\ref{stringschwinger}) arises from the next order in the string loop expansion, 
the cylinder amplitude. 

There is thus  something that is intrinsically stringy about the Schwinger process for strings.  One might speculate that, since we are discussing a transient state of the string theory -- it is a state which is decaying -- it is not an on-shell solution of string theory.  In this sense, the interesting formula (\ref{stringschwinger}) could well be a simple probe of off-shell string theory.

The weak field limit of equation  (\ref{stringschwinger}) was first discussed by Burgess \cite{Burgess1987}.  
Bachas and Porrati \cite{Bachas1992} derived the full expression by finding an operator solution of the string sigma model with an electric field.  The solution was confirmed   using the boundary state technique \cite{Billo1997}.  In this paper, we shall discuss how it can also be obtained by integrating the functional integral for the bosonic string sigma model which describes the appropriate configuration of open strings,  the cylinder amplitude\footnote{
Throughout this paper, we will use the symbols
$\Re $ and $\Im $ for the real and imaginary parts, respectively.
}

\begin{equation} \label{cylinderamp}
\Gamma_{\rm string} =\frac{2}{V} \Im \int_0^\infty \frac{dT}{2T}\int \left[ dX^\mu(\sigma,\tau) \right]
~[{\rm ghost}]~ e^{-S[X,T]}
\end{equation}
where the Polyakov action in the conformal gauge is 
\begin{align}
S=\frac{1}{4\pi\alpha'}\int_0^1d\tau\int_0^1d\sigma
\left[T\dot X_\mu(\sigma,\tau)^2+\frac{1}{T}{X_\mu'}(\sigma,\tau)^2\right] ~~~~~~~~~~~~~~~~~~~~~~~~~~~~~~~~~~~
  \nonumber \\
~~~~~~~~~~~~~~~~~-\frac{E}{2}\int_0^1d\tau \left[X_0(0,\tau)\dot X_1(0,\tau) -
X_1(0,\tau)\dot X_0(0,\tau)\right]
\label{stringactiona}
\end{align}
and the metrics of both spacetime and the string worldsheet are Euclidean. As usual, we denote $\dot x\equiv\frac{\partial}{\partial\tau}x$ and $x'\equiv\frac{\partial}{\partial\sigma}x$.
Here, $T$ is the modular parameter of the cylinder. 
The factor of $1/2$ in the measure reflects the symmetry under time reversal on the annular worldsheet. 
(For the charged scalar particle treated in \cite{Gordon2015,Gordon2016} this factor is absent as the scalar field is complex). The ghost determinant is (see subsection \ref{ghosts})
\begin{align}\label{ghostdet}
	[{\rm ghost}]~ =~\det \left[ -\frac{1}{T}\partial_\sigma^2-T\partial_\tau^2
\right]
\end{align}

In appendix \ref{gaussianintegral}, we will review how the imaginary part of the cylinder amplitude can be computed, and equation (\ref{stringschwinger})
obtained directly by first performing the Gaussian
integral over the embedding coordinates of the string and then finding the imaginary contributions of some poles 
in the modular parameter integral.  
 We can consider this a confirmation of equation (\ref{stringschwinger}) which, as we have discussed above, was found by other techniques.  
 One interesting point is that, when zeta-function regularization is used in order to define the various infinite products and summations which are encountered in taking the functional integral, the result turns out to reproduce  (\ref{stringschwinger}) in every detail, including the overall normalization.   

 Here, we wish to emphasize an alternative approach which, at the outset, appears less efficient.  It is a true semiclassical computation of the partition function where we begin with a classical instanton solution of the equations of motion which are obtained by treating both the embedding coordinates and the modular parameter $T$ as dynamical variables.   The action evaluated on such classical instantons has already been seen to produce the large $m_S^2$ limit of the exponent in equation  (\ref{stringschwinger})  \cite{Schubert2010}.  We will then perform a detailed analysis of the fluctuations about the classical solution, using the Gelfand-Yaglom approach to computing functional determinants.  We will find that, with zeta-function regularization, we produce the full expression in equation  (\ref{stringschwinger}), complete with the prefactors.    This suggests that this semi-classical calculation is giving us the exact result.  We then fashion a proof that the functional integral in an instanton sector is indeed given exactly by the semi-classical, WKB limit.   This proceeds by identifying an interesting nilpotent fermionic transformation of the dynamical variables which is a symmetry of the functional integral. This symmetry is fermionic in that it uses the Fadeev-Popov ghost variables which appear due to a certain gauge fixing, but it differs from the usual BRST supersymmetry. Then we show that the interaction terms in the action in the multi-instanton sector, as well as being closed forms are also exact forms  and can therefore be deformed to zero.   The WKB approximation then gives the exact result.
  
  Although our calculation does not provide any new information beyond equation  (\ref{stringschwinger}) which is already known, we consider it worth presenting nonetheless.  It is one of the few explicit examples of localization of a functional integral and the mechanism for this localization is interesting and new.   It becomes one of a short list of instanton computations that can be done exactly and the sum over all instanton sectors indeed reproduces  (\ref{stringschwinger}), which confirms its validity from a fourth point of view.   
 
 Also, this semiclassical approach can be a starting point for other interesting calculations where the other approaches do not work, for example, where the spacetime
of the D-brane world-volume is curved or where the electromagnetic field is not constant and the integral over embedding coordinates is not Gaussian.  Then the perturbative approach which we espouse would be the conventional starting point.   Although we shall not explore these issues here, it could be that our result -- the knowledge that the flat space, constant field limit is WKB exact -- would be of value in understanding corrections to that limit when spacetime is not flat and gauge fields are not constant or where the gauge fields are dynamical.\footnote{Some exact results for pair-production in non-constant background fields have been obtained in the case of QED; see \cite{Ilderton2014} and references therein. For the QED analog of the present analysis, including of our localization calculation, see \cite{Gordon2015}.}   
 
The rest of the paper is structured as follows. In Section \ref{instanton}, we perform a semi-classical  computation of the cylinder amplitude in the bosonic open string sigma model.   We discuss the instanton solution and we consider the fluctuations of the dynamical variables of the sigma model about the classical solution in the Gaussian approximation. 
 We find that these, subsequently summed over all instanton numbers, gives the exact result, the formula in equation  (\ref{stringschwinger}).   In Section \ref{exact} we shall find a fermionic symmetry of the theory in the multi-instanton sector.   We then use this fermionic symmetry to demonstrate that all perturbative corrections to the  WKB limit indeed cancel so that the WKB approximation is exact.

In Appendix \ref{toymodel} we examine, as a toy model, a simple integral which has features similar to the string functional integral that we evaluate, and which illustrates the technique of evaluating the instanton amplitude, including exactness of the WKB approximation. The supersymmetry that is identified and which we use there is a 
 close analog of one that we have previously found for the 
 particle in an electric field in \cite{Gordon2015,Gordon2016} and the string in an electric field in this paper.   
In appendix \ref{gaussianintegral} we evaluate the imaginary part of the cylinder amplitude \eqref{cylinderamp} by first integrating over the string embedding coordinates and subsequently picking up the imaginary contributions of poles in the integral over the modular parameter $T$.
In appendix \ref{neumannmodeexpansion} we carry out the computation of section \ref{instanton} using an explicit mode expansion instead of the Gelfand-Yaglom method.
In subsequent appendices, we collect some properties of the Dedekind eta-, Jacobi theta and Riemann zeta-functions and their transformations which we make use of in the computations in section \ref{instanton}.

\section{Semiclassical evaluation of the cylinder amplitude} \label{instanton}
Now let us consider the case of a bosonic string in an electric field.  The scenario we are interested in has an
open string that is suspended between two parallel D-branes which both have flat geometry.  We turn on a constant
U(1) electric field in one of the D-branes.
We are interested in the amplitude for the creation of pairs of charged states of the string by the same tunnelling process
as the Schwinger process for particles.  We will   discuss
a semiclassical computation which takes into account worldsheet instantons.  

The instanton solution of the open string theory has already been found and shown to produce the classical limit 
of the amplitude \cite{Schubert2010}. In this section, we shall expand on that calculation. In particular, we will include fluctuations about the classical instanton.  

To evaluate the amplitude for the  Schwinger process for charged strings in an electric field, we shall look for
an imaginary part of the cylinder amplitude.  We begin with the   amplitude in the conformal gauge appearing
in equations (\ref{cylinderamp}), (\ref{stringactiona}) and (\ref{ghosts}).  For the coordinates which are affected by the electric field, it is convenient to use the complex combination
\begin{align}
z(\sigma,\tau)=\frac{1}{\sqrt{2}}\left[X_0(\sigma,\tau)+iX_1(\sigma,\tau)\right]
\end{align}
In this notation, the action becomes
\begin{align}
S=\frac{1}{2\pi\alpha'}\int_0^1d\tau\int_0^1d\sigma \left[ T|\dot z(\sigma,\tau)|^2+\frac{1}{T}|z'(\sigma,\tau)|^2\right]
 +iE\int_0^1 d\tau \bar z(0,\tau)\dot z(0,\tau) \nonumber \\+
\sum_{a=2}^{D}\frac{1}{4\pi\alpha'}\int_0^1d\tau\int_0^1d\sigma
\left[T\dot x_a(\sigma,\tau)^2+\frac{1}{T}{x_a'(\sigma,\tau)}^2\right] \nonumber \\
+\sum_{A=D+1}^{25}\frac{1}{4\pi\alpha'}\int_0^1d\tau\int_0^1d\sigma
\left[T\dot x_A(\sigma,\tau)^2+\frac{1}{T}{x_A'(\sigma,\tau)}^2\right] 
\label{stringaction}
\end{align}
where we label the coordinates which have Neumann boundary conditions as 
$x_a(\sigma,\tau)$, with $a=0,....,D$ and those which have Dirichlet boundary conditions as  $x_A(\sigma,\tau)$
with $A=D+1,...,25$.
The D-brane  is flat and infinite, filling the spacetime coordinates $x_0,...,x_D$.  
 The boundary conditions are periodic in 
worldsheet time, $X_\mu(\sigma, \tau+1)=X_\mu(\sigma,\tau) $ and 
\begin{align}
 z'(\tau,\sigma=1)&=0~~,~~
  z'(\tau,\sigma=0)=2\pi \alpha' iET\dot z(\tau,\sigma=0) \label{efbc1}\\
   x_a'(\tau,\sigma=1) &=0~~,~~ x_a'(\tau, \sigma=0)=0 ~~,~~a=2,...,D  \label{efbc2} \\
 x_A(\tau,\sigma=1)&=d_A   ~~,~~x_A(\tau,\sigma=0)=0~~,~~ A=D+1,...,25   \label{efbc3}
\end{align} 
As usual, the presence of the electric field, which will not appear in the equations of motion, 
is in the boundary condition (\ref{efbc1}).

 
 \subsection{Worldsheet instantons}

We will treat the string coordinates $z(\sigma,\tau),x_a(\sigma,\tau),x_A(\sigma,\tau)$ and the modular parameter $T$ as dynamical variables. We will begin by finding the saddle points of the integrand in the functional  integral by solving  the classical equations of motion. Those equations  are obtained by applying the variational principle to the action (\ref{stringaction}). The 
equations  are
\begin{align}
&z''(\sigma,\tau)+T^2\ddot z(\sigma,\tau)=0   \\
&x''(\sigma,\tau)+T^2\ddot x(\sigma,\tau)=0\\
&\int d\tau d\sigma\left[T^2\left( |\dot z(\sigma,\tau)|^2+\frac{1}{2} \dot x(\sigma,\tau)^2 \right)-
\left(|z'(\sigma,\tau)|^2+\frac{1}{2}{ x'(\sigma,\tau)}^2 \right)\right]=0
\end{align} 
The solutions must also obey the boundary conditions  
(\ref{efbc1})-(\ref{efbc3}). With these boundary conditions, the solutions of the above equations are
\begin{subequations} \label{classicalsolution}
\begin{align}
&z_0(\sigma,\tau)= \frac{1}{\sqrt{2}}~|\vec d|~e^{2\pi ik\tau}\frac{\cosh(2\pi\alpha' {\mathcal E}(\sigma-1)) }{2\pi\alpha' {\mathcal E}}~,~k=1,2,\ldots   \label{cl1} \\
&x_{0a}(\sigma,\tau)=0 
~,~x_{0A}(\sigma,\tau)= d_A\sigma  ~,~
T_0 =\frac{2\pi\alpha' {\mathcal E}}{2\pi k}    \label{cl2}
\end{align}
\end{subequations}
where  $2\pi\alpha'{\mathcal E}=  {\rm arctanh} 2\pi\alpha' E $. 
The solution is a worldsheet with cylindrical topology which intersects the two parallel D-branes between which it is 
suspended on circles.  The 
positive integer $k$ is the instanton number.  It is the number of times that the embedding wraps the cylinder.   

The circle  which the endpoint of the open string traces on the D-brane with the electric field can be thought of as a cyclotron orbit.  In Euclidean space, an electric field behaves as a magnetic field and the charged particle residing at the end of the string follows a cyclotron orbit. The radius of the circle is related to the magnitude of the electric field and 
the separation of the D-branes.  Since the minimum (classical)  mass of the classical string state is given by 
the string tension times the string length, $$ \tilde m_0=\sqrt{ \left(\frac{|\vec d| }{2\pi\alpha'}\right)^2-\frac{1}{\alpha'} }
\approx  \frac{|\vec d| }{2\pi\alpha'} $$
the radius of the orbit is 
\begin{align}
{\rm radius}_E~=~ m_0\frac{\cosh(2\pi\alpha' {\mathcal E}) }{ {\mathcal E}}= \frac{1}{\sqrt{1-\frac{E^2}{(2\pi\alpha')^2}}}
\frac{m_0}{  {\rm arctanh}(2\pi\alpha' E)/2\pi\alpha'}
\end{align}
which is  equal to the   cyclotron radius of a relativistic particle, $\frac{m_0}{E}$ when $E<<2\pi\alpha'$ but gets
very large as $E$ approaches the string scale $2\pi\alpha'$, going to infinity at the critical field. The string endpoint on the D-brane with no electric field also gets dragged in a circle which  has a smaller radius $${\rm radius}_0= m_0\frac{1 }{ {\mathcal E}}=
\frac{m_0}{  {\rm arctanh}(2\pi\alpha' E)/2\pi\alpha'}~=~\sqrt{1-\frac{E^2}{(2\pi\alpha')^2}}
\cdot{\rm radius}_E
$$
The classical action in this $k$-instanton sector is given by
\begin{align} \label{Sclassical}
\boxed{~S_{\rm classical} = \frac{\pi k}{{\mathcal E}} \frac{\vec d^2}{(2\pi\alpha')^2} =  \frac{\pi   m_0^2}{{\mathcal E}}k ~}
\end{align}
which matches the first, dominant term in the exponent in the string amplitude in equation (\ref{stringschwinger}). This
formula is known from previous work \cite{Schubert2010}.
It also approaches the classical action for the cyclotron orbit of a relativistic particle, $\frac{ \pi  m_0^2}{E}k $,  
when $E<<\frac{1}{2\pi\alpha'}$.  When $E$ is of order the string scale, the result is much smaller than that for a particle, 
going to zero at the critical  $E\to\frac{1}{ 2\pi\alpha'}$.

For the remaining sections we define
\begin{equation}
	\varepsilon \equiv 2\alpha' \mathcal{E}
\end{equation}
to bring our formulae in line with the notation of e.g. \cite{Bachas1992,Abouelsaood1987,Schubert2010}.

\subsection{Fluctuations about the instanton}

We now expand the dynamical variables about the classical solution obeying these boundary conditions as
\begin{align} \label{classical}
& z(\sigma,\tau) = z_0(\sigma,\tau) +\delta z(\sigma,\tau) ,~~~~ x_{a}(\sigma,\tau)=\delta x_a(\sigma,\tau) , \nonumber \\
& x_{A}(\sigma,\tau)= d_A\sigma  +\delta x_A(\sigma,\tau) ,~~~~ T = T_0 +\delta T  , 
\end{align}
where the boundary conditions for the fluctuations are
\begin{subequations} \label{boundaryconditions}
\begin{align} 
 \delta z'(\tau,\sigma=1)&=0~~,~~
 \delta  z'(\tau,\sigma=0)=2\pi \alpha' iET \, \delta \dot z(\tau,\sigma=0) \label{efbc10}\\
 \delta  x_a'(\tau,\sigma=1) &=0~~,~~\delta x_a'(\tau, \sigma=0)=0 ~~,~~a=2,...,D  \label{efbc20} \\
\delta  x_A(\tau,\sigma=1)&=0   ~~,~~\delta x_A(\tau,\sigma=0)=0~~,~~ A=D+1,...,25   \label{efbc30}
\end{align} 
\end{subequations}
The expansion of the action to quadratic order in the fluctuations is 
\begin{eqnarray} \label{Sfluc}
	S &=& k m_0^2 \frac{2\pi \alpha' }{\varepsilon} + \frac{1}{2\pi \alpha' }\frac{\d T^2}{T_0^3} \int_0^1 d\sigma d\tau \left[  |z_0^{\prime} |^2 + \frac{1}{2} x_{0A}'^2 \right] +\left. iE\int_0^1d\tau \delta\bar z\delta\dot z\right|_{\sigma=0} \nonumber \\
 && +\frac{1}{2\pi\alpha'}\frac{\delta T}{T_0^2}\int_0^1 d\sigma d\tau \left[ T_0^2
(\dot{\bar z}_0 \delta\dot z+\delta\dot{\bar z}\dot z_0) 
- ( \bar z_0' \delta  z'+\delta\bar z' z_0')\right] \nonumber \\
&& +\frac{1}{2\pi\alpha'}\int_0^1d\sigma d\tau\left[ T_0|\delta \dot z|^2+\frac{1}{T_0}|\delta z'|^2+
\frac{T_0}{2}\delta\dot x^2+\frac{1}{2T_0}{\delta x'}^2\right] + \ldots
\end{eqnarray}
where $T_0$, $z_0$, $x_{0A}$ are the classical solutions given in (\ref{cl1},\ref{cl2}), and we defined $\delta \vec x \equiv \binom{\delta \vec x_a}{\delta \vec x_A}$. Terms of cubic and higher order in fluctuations will be studied in section \ref{exact}. This stringy fluctuation path integral has a  similar structure to that of the  worldline fluctuation integral mediating the Schwinger effect in scalar QED \cite{Gordon2015}. Accordingly, the mechanism by which it reduces to its semiclassical approximation will largely parallel that of \cite{Gordon2015}, up to some technical complications.

\subsubsection{Structure of fluctuation integral} \label{structureofflucs}

Let us set up the evaluation of the above fluctuation integral with care, as it will streamline the calculation considerably. Ignoring for the moment the second line of \eqref{Sfluc}, we have a quadratic form in each of $\d x_a$, $\d x_A$, $\d z$, with identical fluctuation operator
\begin{equation}
	\hat L = \frac{1}{2\pi\alpha' T_0} \left( -\partial_\sigma^2- T_0^2 \partial_\tau^2 \right)
\end{equation}
but different boundary conditions. The $\tau$-dependence is trivially diagonalized by Fourier transforming, 
\begin{equation}
	\delta z (\sigma,\tau) = \sum_{n=-\infty}^\infty e^{2\pi i n \tau} \delta z_n(\sigma) ,
    \qquad \delta x_{i} (\sigma,\tau) = \sum_{n=-\infty}^\infty e^{2\pi i n \tau} \delta x_{i,n}(\sigma) 
\end{equation}
thereby reducing the 2-$d$ spectral problem to 1-$d$, with Robin, Neumann and Dirichlet boundary conditions for $\d z, \d x_{a}$, and $\d x_{A}$ respectively. Gaussian integration then generates a product over $n$ of functional determinants for the 1-$d$ operator
\begin{equation} \label{Ln}
	\hat L_n = \frac{1}{2\pi\alpha' T_0} \left( -\partial_\sigma^2+ \Omega_n^2 \right), \qquad \Omega_n \equiv 2\pi n T_0.
\end{equation}
Both the determinants themselves and their product over $n$ are formally infinite and will require regularization.

At this stage we note a couple of complications. Firstly,
we have ignored the coupling between $\dT$ and $\d z$ in the second line of \eqref{Sfluc}. $\d T$ couples to an infinite number of $n=k$ modes of $\Re \d z$.\footnote{Unlike the analogous worldline path integral for scalar QED  \cite{Gordon2015}, where $\dT$ coupled only to the single mode $z_0$.} Second is the issue of zero modes. Both $z_0$ and $\dot z_0$ are non-constant eigenfunctions of $\hat L_k$ with zero eigenvalue\footnote{
There is also a constant zero mode for the $0,\ldots,(25$-$D$) directions, which generates a factor of the brane worldvolume.
}.

Let us label the normalized eigenmodes of $\hat L_k$ with Robin boundary conditions, and their respective Fourier coefficients, as follows
\begin{equation}
\begin{array}{rll}
\text{Zero modes}:& \qquad  \hat z_0(\sigma) \equiv \frac{z_0(0,\sigma)}{\modd{z_0}}, &  \qquad \v \qquad  \\
 & \qquad \hat{\dot z}_0(\sigma) \equiv \frac{\dot z_0(0,\sigma)}{\modd{\dot z_0}}, & \qquad \z \qquad \\
\text{Non-zero modes}:& \qquad y_i(\sigma) \quad (i>0),  & \qquad \y_i \qquad
\end{array} 
\end{equation}
For later convenience we introduce the following notation for the inner product and norm, respectively, on the Hilbert space $\mathcal L^2\left( [0,1]\times [0,1] \right)$:
\begin{eqnarray}
	\ev{z|w} &\equiv & \int_0^1 \! d\tau \! \int_0^1 \! d\sigma \, \left[ \bar z(\sigma,\tau) w (\sigma,\tau) + \bar w (\sigma,\tau) z (\sigma,\tau) \right] \\
	\modd{z(\sigma,\tau)} &\equiv & \sqrt{\ev{z|z}}
\end{eqnarray}
Similarly, for the real coordinates $X_\mu$ ($\mu=2,\dots,25$) we define
\begin{eqnarray}
\ev{X|Y} &\equiv & \int_0^1 \! d\tau \! \int_0^1 \! d\sigma \, \vec X(\sigma,\tau) \cdot \vec Y (\sigma,\tau) \\
 \modd{X(\sigma,\tau)} &\equiv &  \sqrt{\ev{X|X}} .
\end{eqnarray}
$\hat{\dot z}_0(\sigma)$ is a genuine zero mode of the quadratic fluctuation action \eqref{Sfluc}, i.e. the action is independent of $\z$. This is a familiar consequence of expanding a $\tau$-translationally invariant action about a $\tau$-dependent solution, $z_0(\sigma,\tau)$. This will have to be gauge-fixed and the Faddeev-Popov jacobian $J_{FP}$ accounted for. We do this in subsection \ref{zeromode}. On the other hand, $z_0(\sigma,\tau)$ couples linearly to $\delta T$, and therefore is not a genuine zero mode of the action.

In fact these comments lead to a key simplifying observation regarding  the combined $\d z_k(\sigma)$, $\dT$ quadratic form. This part of the fluctuation action \eqref{Sfluc} can be written, schematically, as
\begin{equation} \label{Sdzdt}
	S_{\d z,\dT} \propto \frac{1}{2}a \dT^2+\dT (b \,\v+\vec c \cdot \vec\y) + \frac{1}{2}\vec \y^T D \,\vec \y
\end{equation}
where the values of the constants $a$, $b$, $\vec c$ and matrix $D$ can be read off from \eqref{Sfluc}. Now this quadratic form is not positive definite, but must be defined by analytic continuation, $\v\rightarrow \pm i \v$. It is in this manner that the path integral obtains an imaginary part, and thus a non-zero tunneling probability. The advertised simplification is that all dependence on both $\vec c$ and $a$ cancels out once we integrate out $\dT$, $\v$, and $\vec\y$:
\begin{equation} \label{gauss}
	\int d\dT d\v \prod_{i=1}^N d\y_i e^{-S_{\d z,\dT}} = \pm\sqrt{-1} \frac{(2\pi)^{n+1}}{b \,\det D}
\end{equation}
We have left implicit the limit $N\rightarrow \infty$. The upshot is that we can as well make the replacement $\d z(\sigma,\tau) \rightarrow \hat z_0(\sigma, \tau) \,\v$ in the second line of \eqref{Sfluc}, to determine the tachyonic contribution denoted $b$ in \eqref{gauss}, and we simply omit the zero eigenvalue in our evaluation of $\det \hat L_k$.

The semiclassical approximation to \eqref{cylinderamp} can then be summarized by the following expression:
\begin{multline} \label{semiclassical}
		\Gamma_{\rm string} \simeq \frac{2}{V} \Im  \frac{V}{2T_0}\left(\prod\sqrt{2\pi}\right) e^{-S_{\rm classical}}
~\left.[{\rm ghost}]\right|_{T_0} \left[ \left( \det_R' \hat L_0 \right)\left( \det_R' \hat L_k\right) \prod_{n\neq 0,k} \det_R \hat L_n\right]^{-1} \\
\left[ \det_N' \hat L_0 \nprod \det_N \hat L_n \right]^{-\frac{D-1}{2}}
\left[ \prod_{n=-\infty}^\infty \det_D \hat L_n \right]^{\frac{D-25}{2}} \cdot J_{FP}^{classical}\cdot
(\rm{tachyon})
\end{multline}
Here $J_{FP}\cl$ is the leading-order (in fluctuations) part of the Faddeev-Popov determinant. By ``tachyon'' we mean the imaginary contribution of the $\dT$,$\d z$ form (this is described above and indicated by $\pm i/b$ in equation \eqref{gauss}). The expression $(\prod\sqrt{2\pi})$ indicates the Gaussian integration normalization, and will be carefully accounted for later. The operator $\hat L_n$ is defined in \eqref{Ln}. The subscripts $R(obin)$, $N(eumann)$ and $D(irichlet)$ refer to the relevant boundary conditions:
\begin{subequations}
\begin{align} 
R: \qquad  \delta z_n'(\sigma=1)&=0~~,~~
 \delta  z_n'(\sigma=0) + 2\pi \alpha'E \Omega_n T_0 \, \delta z_n(\sigma=0) = 0 \\
N: \qquad \delta  x_a'(\sigma=1) &=0~~,~~\delta x_a'(\sigma=0)=0 ~~,~~a=2,...,D  \\
D: \qquad \delta  x_A(\sigma=1)&=0   ~~,~~\delta x_A(\sigma=0)=0~~,~~ A=D+1,...,25  
\end{align} 
\end{subequations}
We now proceed to determine each of the above components. Evaluation of the functional determinants is straightforward if one knows the eigenvalues. For the Neumann/Dirichlet cases this is not a problem, but for Robin boundary conditions the eigenvalues are not explicitly known, being determined by a transcendental equation. One option then is to expand the fluctuations $\delta z_n$ in modes satisfying incorrect boundary conditions, say $\d z '=0$. The quadratic form will not be diagonal in this basis, so that more work is required to evaluate its determinant, but we show in appendix \ref{neumannmodeexpansion} that this can be done in detail and it obtains the same result as the one which we derive below.

A more elegant approach which does not rely on knowledge of the eigenvalues, and which will generalize readily to more complicated setups (e.g. non-constant background fields), is the method of Gelfand-Yaglom. We employ this method in subsection \ref{GY} to evaluate all determinants in \eqref{semiclassical}.

\subsubsection{Zero mode} \label{zeromode}
Since $Im [\delta z]$ does not couple to $\delta T$, in this case the zero eigenvalue of $\hat L_k$ corresponds to a genuine zero-mode of the quadratic action. This is expected, and results from proper-time-translation invariance and the fact that the instanton depends on the world-sheet time. Denoting the gauge transformation parameter by $t$ as follows
\begin{equation}
	z_t (\sigma,\tau) = z (\sigma,\tau+t)
\end{equation}
one has that $\dot{z}_{0,t}(\tau)$ is a zero mode of the quadratic action:
\begin{equation}
	\left.  \frac{\delta^2 L}{\delta z \delta \bar{z}} \right|_{z_{0,t}} \frac{d}{dt} z_{0,t} = \left. \frac{d}{dt} \left(\frac{\delta L}{\delta \bar{z}}\right) \right|_{z_{0,t}} = 0
\end{equation}
We gauge fix by introducing a collective coordinate. The Faddeev-Popov trick begins by introducing unity into the path integral in the form
\begin{align}\label{fp}
	1=\frac{1}{\omega}\int_0^1 \! dt ~ \delta \left( g(t)  \right) \cdot \frac{d}{dt} g(t)
\end{align}
The gauge-fixing function $g(t)$ is chosen so as to render the integration over the zero mode well-defined. The ``Gribov'' factor $\omega$ is the number solutions of $g(t) = 0$ in the interval $0<t<1$. We choose
\begin{equation} \label{gaugefixfn}
	g(t) = \frac{1}{\modd{\dot z_0}} \ev{z_t|\dot z_0} ,
\end{equation}
for which 
\begin{equation}
	\omega = 1/2k .
\end{equation}
The classical solution $z_0(\sigma, \tau)$ was defined in \eqref{classicalsolution}. Then the time translation symmetry of the path integral is used to translate the argument $t$ to zero. This procedure then amounts to inserting 
\begin{equation}
	\delta(\z)\, \frac{\dot g(0)}{2k}
\end{equation}
where
\begin{eqnarray}
	\dot g(0) \equiv \left. \deriv{}{t} g(t) \right|_{t=0} &=& \frac{1}{\modd{\dot z_0}} \ev{\dot z| \dot z_0}\\
    &=& \modd{\dot z_0} + \frac{1}{\modd{\dot z_0}} \ev{\d \dot z| \dot z_0}
\end{eqnarray}
This allows the integration over $\z$ to be done using the delta function, and leaves a Faddeev-Popov jacobian whose classical contribution is the first, constant term above. The second term  is a correction which we ignore in the present semi-classical computation. The net result then is the insertion of
\begin{equation}
	J_{FP}^{classical} = \pi \modd{z_0}
\end{equation}

\subsubsection{Ghosts} \label{ghosts}
For completeness we briefly review here the contribution of reparametrization ghosts to the path integral. These arise from gauge-fixing of the Diffeomorphism $\otimes$ Weyl symmetry on the worldsheet, as explained in detail in e.g. \cite{Polchinski2007}. An arbitrary, infinitesimal such symmetry variation is given by
\begin{equation}
	\delta g_{ab} = 2\omega g_{ab}-\nabla_a\delta \sigma_b -\nabla_b\delta \sigma_a
\end{equation}
where $(\sigma^0,\sigma^1)\equiv (\sigma,\tau)$, $g_{ab}$ is the worldsheet metric, and $\omega$ parametrizes an infinitesimal Weyl transformation. The Faddeev-Popov procedure applied to this invariance leads to the following ghost action in conformal gauge:\footnote{
In this subsection we use the following notation:
\begin{eqnarray*}
	z&=& \sigma + i \tau\\
    \partial_z, \partial_{\bar z} &=& \frac{1}{2} (\partial_\sigma \mp i\partial_\tau) \\
    d^2z &=& 2d\sigma d\tau
\end{eqnarray*}
}
\begin{eqnarray}
	S_{gh} &=& \frac{1}{2} \int \! d^2z \left( b\partial_{\bar{z}} c + \tilde{b} \partial_z \tilde{c} \right) \\
    &\equiv& \frac{1}{2} \int_0^{1/T} \! d\tau \int_0^1 \! d\sigma \left\{ b(\partial_\sigma +i\partial_\tau)c + \tilde{b}(\partial_\sigma -i\partial_\tau)\tilde{c}\right\}
\end{eqnarray}
The integration limit $1/T$ follows from our definition of $T$ in \eqref{stringaction} (usually the modular parameter is taken as the reciprocal of this). $b$, $\tilde b$, $c$ and $\tilde c$ are ghost fields whose boundary conditions
\begin{equation}
	c=\tilde{c}, \quad b=\tilde{b}, \quad \text{on boundary},
\end{equation}
are inherited from the worldsheet reparametrizations.
These conditions are easily implemented using the so-called doubling trick. Define $B(\sigma,\tau)$, $C(\sigma,\tau)$ on the doubled domain $\sigma\in[-1,1]$ by
\begin{eqnarray}
	C(0<\sigma<1)\equiv c(\sigma), \quad C(-1<\sigma<0) \equiv \tilde{c}(-\sigma) \\
    B(0<\sigma<1)\equiv b(\sigma), \quad B(-1<\sigma<0) \equiv \tilde{b}(-\sigma)
\end{eqnarray}
with periodic boundary conditions on $B$, $C$. The ghost action is then
\begin{equation}
	S_{gh} = \int_0^{1/T} \!\! d\tau \int_{-1}^1 \!\! d\sigma \, B(\sigma,\tau) \partial_{\bar{z}} C(\sigma,\tau)
\end{equation}
and the expansion
\begin{equation}
	B(\sigma,\tau) = \sum_{m=-\infty}^\infty \sum_{n=-\infty}^\infty e^{2\pi i n \tau T + \pi i m \sigma} B_{mn} \, , \qquad C(\sigma,\tau) = \sum_{m=-\infty}^\infty \sum_{n=-\infty}^\infty e^{2\pi i n \tau T + \pi i m \sigma} C_{mn}
\end{equation}
leads to (we exclude the simultaneous zero mode $m=n=0$)
\begin{eqnarray}
	[\text{ghost}] &=& \left(\prod_{n\neq 0} (2\pi n)\right) \prod_n \prod_{m\neq 0} \left[ 2\pi n-i\pi m /T\right] \\
    &=& \eta^2(i/2T)
\end{eqnarray}
As usual we have used $\zeta$-regularization - see appendix \ref{zeta}. For the semiclassical approximation we approximate $T\simeq T_0$ to get
\begin{equation}
	\left. [\text{ghost}]\right|_{T_0} = \eta^{2}(ik/\epsilon) .
\end{equation}

\subsubsection{Tachyon}
As described in section \ref{structureofflucs}, the variables $\dT$ and $\v$ couple in the following quadratic form
\begin{equation}
	\begin{matrix}
	(\dT, & \v) \\
     & 
	\end{matrix}
	\begin{pmatrix}
	a & b/2 \\
    b/2 & 0
	\end{pmatrix} \binom{\dT}{\v}
\end{equation}
and we can ignore the coupling of $\d T$ to the other modes.
$b$ is simply the coefficient of $\v \cdot \dT$ obtained by substituting $\d z(\sigma,\tau) \rightarrow \hat z_0(\sigma,\tau)\, \v $ in \eqref{Sfluc}, and we find
\begin{equation}
	(\text{tachyon}) = \pm i \frac{\pi}{b} = \pm i \frac{\alpha'}{k^2} \frac{\modd{z_0}}{R^2}
\end{equation}
where $R \equiv \frac{d}{\pe}$.

\subsection{Determinants \`a la Gelfand-Yaglom from contour integration} \label{GY}

The technique of evaluating functional determinants by relating them to the solution of an initial value problem is originally due to Gelfand and Yaglom \cite{Gelfand1960}. It has since been extended in various directions, including more general boundary conditions, operators with zero modes and partial differential operators, \cite{Forman1987,McKane1995,Kirsten2003,Kirsten2004,Kirsten2008,Kirsten2010}. See \cite{Dunne2008} for a review. Here we briefly review an elegant and simple derivation based on contour integration \cite{Kirsten2003}, for ordinary differential operators with quite general boundary conditions.

Given an ordinary differential operator $\hat L$ and boundary conditions on $[0,1] \subset \mathbb{R}$ we can associate to the spectral problem $\hat L \psi_i(x) = \lambda_i \psi_i (x)$ a generalized zeta function
\begin{equation} \label{zetasum}
	\zeta(s) \equiv \sum_{i=1}^\infty \lambda_i^{-s} \, .
\end{equation}
For simplicity we will assume positive eigenvalues, $\lambda_i >0$. The above sum is only convergent for sufficiently large $s$; however, $\zeta(s)$ can be analytically continued to give a meromorphic function on $\mathbb{C}$. The determinant of $\hat L$, which nominally is given by the divergent infinite product $\prod_i \lambda_{i>0}$, is then \emph{defined} in zeta function regularization as
\begin{equation} \label{detL}
	\det \hat L \equiv \exp (-\zeta'(0))
\end{equation}
Now suppose we have a function $\mathcal U(\lambda)$ with  zeros at precisely the eigenvalues $\lambda = \lambda_i$ of $\hat L$ (without necessarily knowing the values of these eigenvalues). We will come to the problem of constructing such a function presently. Then the logarithmic derivative of $\mathcal U(\lambda)$ has poles of unit residue at each $\lambda_i$, and we therefore have the following contour integral representation of the sum \eqref{zetasum}: 
\begin{equation} \label{zetacontour}
	\zeta(s) = \frac{1}{2\pi i} \int_\gamma \!d\lambda \, \lambda^{-s} \frac{d}{d \lambda} \ln \mathcal U(\lambda)
\end{equation}
The contour $\gamma$ encloses all poles $\lambda_i \in \mathbb{R}^+$ in a counterclockwise sense. Deforming it to the negative real axis we then obtain
\begin{equation}
	\zeta(s) = \frac{\sin \pi s}{\pi} \int_0^\infty \!d\lambda \, \lambda^{-s} \frac{d}{d \lambda} \ln \mathcal U(-\lambda)
\end{equation}
Clearly the above expressions are not convergent for all $s$, (and in particular not for $s=0$). The usual way of handling this is to compute the ratio of determinants (with respect to the same boundary conditions) of $\hat L$ and some simpler normalizing operator $\hat L_0$ whose spectrum is known. This improves the convergence; in particular, if the coefficient of the derivative of highest degree in $\hat L_0$ is the same as for $\hat L$, the ratio of their determinants is finite. Since the large-$\lambda$ asymptotics of $\mathcal U$ and $\mathcal U_0$ are the same, the representation
\begin{equation}
	\zeta_1(s)-\zeta_2(s) = \frac{\sin \pi s}{\pi} \int_0^\infty \!d\lambda \, \lambda^{-s} \frac{d}{d \lambda} \ln \frac{\mathcal U_1(-\lambda)}{\mathcal U_2(-\lambda)}
\end{equation}
is valid around $s=0$, and one has $\det \hat L/\det \hat L_0 = \exp \left\{ -\left[\zeta'(0) - \zeta_1'(0)\right] \right\} $.

For the fluctuation problem in this paper, the main difficulty stems from the Robin boundary conditions. Even for the ``massless'' operator $-\partial^2$ we do not know the spectrum explicitly. Therefore we will study the ``absolute'' determinant directly, performing an explicit analytic continuation.

The idea then is to improve the large-$\lambda$ behavior. By splitting the integration range and subtracting off the leading asymptotic form
\begin{equation}
	\mathcal U(-\lambda) \overset{\lambda\rightarrow\infty}{\sim} \mathcal U_\infty(-\lambda)
\end{equation}
of $\mathcal U$ we have \cite{Kirsten2010,Kirsten2008}
\begin{equation}
	\zeta(s) = \zeta_{fin}(s) + \zeta_{\infty}(s)
\end{equation}
where the finite and asymptotic contributions are respectively
\begin{eqnarray}
	\zeta_{fin}(s) &=& \frac{\sin \pi s}{\pi} \left( \int_0^1 \!d\lambda \, \lambda^{-s} \frac{d}{d \lambda} \ln \mathcal U(-\lambda) + \int_1^\infty \!d\lambda \, \lambda^{-s} \frac{d}{d \lambda} \ln  \frac{\mathcal U(-\lambda)}{\mathcal{U}_\infty(-\lambda)} \right) \label{finitezeta}\\
    \zeta_\infty(s) &=& \frac{\sin \pi s}{\pi} \int_1^\infty \!d\lambda \, \lambda^{-s} \frac{d}{d \lambda} \ln \mathcal U_\infty(-\lambda) \label{asymptoticzeta}
\end{eqnarray}
Since $\mathcal U_\infty$ is a much simpler function, it will be possible to evaluate the integral \eqref{asymptoticzeta} explicitly (assuming large $s$), and the resulting meromorphic expression taken to define the analytic continuation of $\zeta_\infty$ by allowing $s\in \mathbb C$. Meanwhile $\zeta_{fin}(s)$ is now well-defined at $s=0$, and we have
\begin{equation}
	\zeta_{fin}'(0) = -\ln \left[ \frac{\mathcal U(0) }{ \mathcal U_\infty (-1)} \right]
\end{equation}
In conclusion,
\begin{equation} \label{GYdetL}
	\det \hat L = \mathcal U(0) \, \mathcal U_\infty^{-1} (-1) e^{-\zeta_\infty'(0)}
\end{equation}

At this stage the problem boils down to constructing a function $\mathcal U(\lambda)$ with the desired arrangement of zeros. To this end, consider again the eigenvalue equation
\begin{equation} \label{ivproblem}
	\left( \hat L - \lambda \right) u_\l(x) = 0 
\end{equation}
with associated boundary conditions expressed in the general form
\begin{equation} \label{matrixbcs}
	M \binom{u_\l(0)}{u^\prime_\l(0)} + N \binom{u_\l(1)}{u^\prime_\l(1)} = \binom{0}{0}, 
\end{equation} 
for some constant matrices $M$, $N$. Suppose for now, however, we instead impose some arbitrary initial ($x=0$) conditions on \eqref{ivproblem}, leaving the right-hand boundary condition free. This uniquely fixes two independent solutions\footnote{assuming $\hat L$ is of second order} $u_\l^{(1,2)}(x)$ of \eqref{ivproblem}, from which we can construct a general solution $\mathfrak u_\l(x) = \alpha u_\l^{(1)}(x) + \beta u_\l^{(2)}(x)$. Such a solution evaluated at the boundary, $x=1$, defines a function of $\lambda$. It is then straightforward to show that the condition for the existence of a solution $\mathfrak u_\l(x)$ satisfying the  boundary conditions \eqref{matrixbcs} is \cite{Kirsten2003}
\begin{equation}
	\det \left( M + N H_\l(1) H_\l^{-1}(0) \right) = 0
\end{equation}
where $H_\l(x)$ is the fundamental matrix defined by
\begin{equation} \label{fundamentalmatrix}
	H_\l(x) \equiv 
    \begin{pmatrix} 
    u_\l^{(1)}(x) & u_\l^{(2)}(x) \\
    u_\l^{\prime(1)}(x) & u_\l^{\prime(2)}(x)
    \end{pmatrix}.
\end{equation}
It proves convenient to impose the particular initial condition
\begin{equation} \label{fundamentalIC}
	H_\l(0) = \begin{pmatrix}
	1 & 0 \\ 
    0 & 1
	\end{pmatrix}
\end{equation}
Then the function we are after is given by
\begin{equation}
	\mathcal U(\lambda) = \det \left( M + N H_\l(1) \right)
\end{equation}

\subsubsection*{Exclusion of zero eigenvalue}

When $\hat L$ has a zero mode, the usual approach is to introduce an ad hoc regulator and divide out the pseudo-zero-eigenvalue by hand. A more rigorous, regulator-independent approach detailed in \cite{Kirsten2003,Kirsten2004} involves a slight modification to the above zeta-function computation. We briefly summarize their results, as pertaining to the calculation in this paper. 

The required determinant $\det^\prime \hat L$ with zero eigenvalue excluded follows from (\ref{detL},\ref{zetacontour}) if we modify $\zeta(s)$ appropriately. Namely, we replace $\mathcal U(\l)$ in \eqref{zetacontour} by a function $f(\l)$ with the same positive zeros, but which is non-zero at the origin. This allows for $\gamma$ to be deformed to $\mathbb{R}^-$ as before, by eliminating the singularity at $0$. Such a function is given by
\begin{equation}
	f(\l) \equiv  -\frac{1}{\l} \det \left( M + N H_\l (1) \right)
\end{equation}
Furthermore, if we define
\begin{multline}
	u_\l (x) \equiv -\left[ m_{12} + n_{11} u^{(2)}(1) + n_{12} u^{\prime(2)}(1) \right] u^{(1)}(x)  \\
    + \left[ m_{11} + n_{11} u^{(1)}(1) + n_{12} u^{\prime(1)} (1) \right] u^{(2)}(x)
\end{multline}
with $u^{(1,2)}(x)$ as before, then we have
\begin{equation}
	f(\l) = - \mathcal{B}  \int \! dx \, u_0(x)^* u_\l (x) 
\end{equation}
where the constant $\mathcal B$ is given by
\begin{equation}
	\mathcal{B} = \frac{n_{12}}{u_0'(1)^*} \quad \text{if } n_{12} \neq 0; \qquad \mathcal{B} = \frac{-n_{22}}{u_0(1)^*} \quad \text{if } n_{22} \neq 0 .
\end{equation}
for the general case of ``separable'' boundary conditions, i.e. $\det M = \det N = 0$. Equation \eqref{GYdetL} is thus replaced with
\begin{equation} \label{detprime}
	\det' \hat L = -\mathcal{B} |y(\sigma)|^2 \,  \frac{e^{-\zeta_\infty'(0)} }{ \mathcal U_\infty (-1) }
\end{equation}
where, in the notation of \cite{Kirsten2003}, $y_1(\sigma) \equiv \lim_{\l\rightarrow 0} u_\l(0)$.

In subsection \ref{GYevaluation} we evaluate several determinants with excluded zero modes using a regulator, but it is satisfying to note that the results are reproduced in each case by the formula \eqref{detprime}.

\subsubsection{Evaluation of the Functional Determinants} \label{GYevaluation}

It is now a simple matter to apply the algorithm of section \ref{GY} to the evaluation of $\det_\mu \hat L_n$, for $(\mu=R,N,D)$. We first normalize the ``kinetic'' term to unity, as is customary. That is, we will compute the determinants of the operator 
\begin{equation}
	\L_n = -\partial_\sigma^2 - \Omega_n^2, \qquad \Omega_n \equiv 2\pi n T_0
\end{equation}
This normalization introduces a factor of $\n$ into the integral form of the zeta function as follows:
\begin{equation}
	\zeta(s) \rightarrow \tilde{\zeta}(s) = \frac{\sin \pi s}{\pi} \int_0^\infty \!d\lambda \, (\n \lambda)^{-s} \partial_\lambda \ln \mathcal U(-\lambda)
\end{equation}
Clearly $\tilde \zeta_{fin}'(0) = \zeta_{fin}'(0)$, but the asymptotic piece gives a non-trivial contribution:
\begin{eqnarray}
	\tilde \zeta_\infty '(0) &=& \left. \frac{d}{d s} \left( \n^{-s} \zeta_\infty(s) \right) \right|_{s=0} \\
    &=& (-\log \n) \zeta_\infty(0) + \zeta_\infty '(0)
\end{eqnarray}
Consequently, the relation between the determinants of $\hat L$ and $\L$ is
\begin{equation}
	\det_\mu \hat L_n = \n^{\zeta^{(\mu)}_\infty(0)} \, \det_\mu \L_n,\qquad \mu = R,N,D
\end{equation}
However, the (Riemann-zeta-regularized) product over all $n\in \mathbb Z$ erases this dependence on $\n$, since $\prod_{n=-\infty}^\infty (const) = 1$, and $\zeta_\infty(s)$ does not depend on $n$.
The only such contribution then is from the excluded pseudo-zero-eigenvalues, so effectively each ``prime'' on a determinant is  accompanied by $\n^{-1}$.

For convenience, we will absorb the conventional powers of $\sqrt{2\pi}$ accompanying each Gaussian integration (and hence eigenvalue) into $\n$, defining
\begin{equation}
	\n \equiv \frac{1}{\tpa T_0}\frac{1}{2\pi} = \frac{k}{2\pi^2 \alpha' \epsilon} .
\end{equation}
With these notations, we write \eqref{semiclassical} as
\begin{multline}
		\Gamma_{\rm string} \simeq \Im \frac{2}{V} \frac{V}{2T_0} e^{-S_{\rm classical}}
~\left.[{\rm ghost}]\right|_{T_0} \left[ \n^{-2}\left( \det_R' \L_0 \right)\left( \det_R' \L_k\right) \prod_{n\neq 0,k} \det_R \L_n\right]^{-1} \\
\left[ \frac{\det_N' \L_0 }{\n} \nprod \det_N \L_n \right]^{-\frac{D-1}{2}}
\left[ \prod_{n=-\infty}^\infty \det_D \L_n \right]^{\frac{D-25}{2}} \cdot J_{FP}\cl\cdot
(\rm{tachyon})
\end{multline}

Without further ado, we let us compute the determinants. The fundamental matrix \eqref{fundamentalmatrix} satisfying \eqref{fundamentalIC} is
\begin{equation}
	H_\l (\sigma) = \begin{pmatrix}
	\cos \left( \sigma \sqrt{\l - \Omega^2} \right) & \frac{ \sin \left( \sigma \sqrt{\l - \Omega^2} \right)}{\sqrt{\l - \Omega^2}} \\
    - \sqrt{\l - \Omega^2} \sin \left( \sigma \sqrt{\l - \Omega^2} \right) & \cos \left( \sigma \sqrt{\l - \Omega^2} \right)
	\end{pmatrix}
\end{equation}

\begin{itemize}
\item \textbf{Robin boundary conditions}

In the notation of \eqref{matrixbcs} we have
\begin{equation}
	M = \begin{pmatrix}
	2\pi \alpha' E \Omega_n & 1 \\ 0 & 0
	\end{pmatrix}
    , \quad
    N = \begin{pmatrix}
    0 & 0 \\ 0 & 1
    \end{pmatrix}
\end{equation}
Thus
\begin{align} 
 \mathcal U(-\lambda) &= 2\pi \alpha' E \Omega_n \cosh \sqrt{\l + \Omega^2} - \sqrt{\l + \Omega^2}  \sinh \sqrt{\l + \Omega^2} ~, \\
 \mathcal U_\infty (-\lambda) &= -\frac{1}{2} \sqrt \l e^{\sqrt \l}\nonumber\\
 \zeta_\infty(s) &= \frac{\sin \pi s}{2\pi} \left( \frac{1}{s} + \frac{1}{s-\frac{1}{2}} \right) ~, \qquad \lim_{s\rightarrow 0} \zeta_\infty'(0) = -1
\end{align} 
Therefore recalling \eqref{GYdetL}
\begin{equation}
	\det \L = \mathcal U(0) \, \mathcal U_\infty^{-1} (-1) e^{-\zeta_\infty'(0)}
\end{equation}
we have
\begin{equation}
	\det_R \L_n = 2\Omega_n \sinh\Omega_n \left[ 1 - 2\pi \alpha' E \coth \Omega_n \right] \label{robindet}
\end{equation}
Using the infinite product formulae derived in appendix \ref{zeta}, we obtain
\begin{equation}
	\nkprod \det_R \L_n = \left[ \frac{(-1)^{k+1} e^{-\pi k \epsilon }}{\cosh^2{\pi \epsilon} } \right]^{-1} \frac{\eta^2(ik/\epsilon) }{2\pe \sinh \pe } \frac{2k^2}{\epsilon^2}
\end{equation}
To handle the zero modes, we choose to modify the operators with a regulator $\delta\ll 1$, leaving the boundary conditions unchanged:

\noindent \underline{$n=k$}: Take
\begin{equation}
	\L \rightarrow \Lt = -\partial_\sigma^2 - \Omega^2(1+\delta)^2
\end{equation}
Note that $\dot z_0$ ($z_0$) is still an eigenfunction, but with eigenvalue $\l_0 = 2 \delta (\pe)^2$, and the determinant is\footnote{
Alternatively, the result \eqref{detprime} gives
$$
\det_R' \L_k = -2\mathcal{B}|y(\sigma)|^2 ,
$$
with $y(\sigma) = \tanh(\pe) \sinh (\Omega_k \sigma) - \cosh(\Omega_k \sigma)$ and $\mathcal{B} = \cosh\pe$, agreeing with \eqref{detLk}.}
\begin{equation} \label{detLk}
	\det_R' \L_k = \lim_{\delta\rightarrow 0} \frac{\det_R \Lt_k}{\tilde \lambda_0}  
    = \frac{2\pe + \sinh 2\pe }{2 \pe}  \sech \pe 
    = \frac{4||z_0||^2}{R_0^2 \cosh\pe }
\end{equation}

\underline{$n=0$}: The boundary conditions reduce to pure Neumann ($A_0=0$). Taking $\Lt_0 = -\partial_\sigma^2 + \delta^2$, the zero mode $y_0(\sigma)=1$ is unchanged but eigenvalue shifts to $\delta^2$, giving
\begin{equation} \label{constantzeromode}
	\det_R' \L_0 = 2
\end{equation}
Therefore the ``Robin'' contribution $ \big[ \n^{-2}  \left( \det_R' \L_0 \right)\left( \det_R' \L_k\right) \prod_{n\neq 0,k} \det_R \L_n \big]^{-1}$ is
\begin{equation}
	\text{Robin} = \n^2 \left[ \frac{(-1)^{k+1} e^{-\pi k \epsilon }}{\cosh^2{\pi \epsilon} } \right] \frac{(2\pe) \sinh \pe}{\eta^2(ik/\epsilon) } \left(\frac{\epsilon}{2k}\right)^2 \,\frac{R^2 \cosh\pe }{2||z_0||^2}
\end{equation}

\item \textbf{Neumann boundary conditions}

The result follows from (\ref{robindet},\ref{constantzeromode}) and the infinite product formulae in appendix \ref{zeta} by setting $A=0$:
\begin{eqnarray}
	\text{Neumann} &=& \left[ \n^{-1} \det_N' \L_0 \nprod \det_N \L_n \right]^{-\frac{D-1}{2}} \\
    &=& \n^{\frac{D-1}{2}} \left[ 2\nprod  2\Omega_n \sinh \Omega_n \right]^{-\frac{D-1}{2}} \\
    &=& \left[\frac{\epsilon}{8\pi^2 \alpha' k} \right]^{\frac{D-1}{2}} \eta^{1-D}(ik/\epsilon) 
\end{eqnarray} 

\item \textbf{Dirichlet boundary conditions}

We have $M = \begin{pmatrix}
 1 & 0\\
 0 & 0   \end{pmatrix}$ and $
 N = \begin{pmatrix}
 0 & 0\\
 1 & 0   \end{pmatrix} $, therefore
\begin{align} 
 \mathcal U(-\lambda) &= \frac{\sinh  \sqrt{\lambda+\Omega_n^2}}{\sqrt{\lambda+\Omega_n^2}}~~,~~
 \mathcal U_\infty (-\lambda) = \frac{e^{\sqrt{\lambda}}}{2\sqrt{\lambda}} \nonumber\\
 \zeta_\infty(s) &= \frac{\sin\pi s}{2\pi}\left( \frac{1}{ s-\frac{1}{2}}-\frac{1}{s} \right) ~~,~~ \lim_{s\rightarrow 0} \zeta_\infty'(0) = -1
\end{align} 
and
\begin{equation}
	\det_D \L_n = \frac{2\sinh\Omega_n}{\Omega_n}.
\end{equation}
The net contribution is thus
\begin{equation}
	\text{Dirichlet} = \prod_{n=-\infty}^\infty \left( \det_D \L_n \right)^{\frac{D-25}{2}} = \left[ \eta(ik/\epsilon) \right]^{D-25}
\end{equation}
\end{itemize}

\subsection{Result}
Gathering all factors, we find that the semi-classical approximation to the annular string partition function with action \eqref{stringaction} (see \eqref{semiclassical}) yields
\begin{equation} \label{finalresult}
	\Gamma^{\text{semicl.}} = \frac{\tanh\pe}{\epsilon} \frac{(-1)^{k+1} e^{-\pi k \epsilon }}{\eta^{24}(ik/\epsilon)}  \left[\frac{\epsilon}{8\pi^2 \alpha'k} \right]^{\frac{D+1}{2}} \, e^{-2\pi \alpha' k m_0^2/\varepsilon} ,
\end{equation}
%
%
where we recall that $m_0 = \frac{d}{2\pi\alpha'}$ and $\epsilon=\frac{1}{\pi} \text{arctanh} (2\pi\alpha' E)$. This is identical to the result (equation (25)) of \cite{Bachas1992}. To see this, it is useful to recall the mass-shell relation of the open string,
\begin{equation}
	m_S^2 = \frac{d^2}{(2\pi\alpha')^2} + \frac{1}{\alpha'}\left(\mathcal{N}-1\right)
\end{equation}
where $\mathcal N$ is the level number, from which one obtains 
\begin{equation}
	\sum_S e^{-(2\pi\alpha') k m_S^2/\varepsilon} = e^{-(2\pi\alpha') k m_0^2/\varepsilon} \eta^{-24}(ik/\varepsilon).
\end{equation}
(Note that in contrast to our setup, theirs allows for both string endpoints to be charged, but is specialized to a spacetime-filling D-brane, $D=25$.)

\newpage

\section{Exactness of semiclassical approximation: Proof by localization}\label{exact}
It is remarkable that our semi-classical computation has produced the \emph{exact} amplitude for pair production. It implies that all higher-order corrections must find a way to cancel. In this section we will study the full path integral and construct a proof that it localizes onto its semi-classical approximation.

There is an analogous localization of the worldline path integral of scalar QED in a constant electric field, which we addressed in \cite{Gordon2015} (and which inspired the present investigation). While the underlying mechanism by which the two path integrals localize is essentially the same, the $2$d worldsheet with boundary does introduce some complications relative to the particle worldline, and in particular the manipulations of this section become significantly more cumbersome.

The cleanest approach proceeds by identifying a fermionic symmetry of the gauge-fixed action, mixing ghost and bosonic variables (but distinct from the usual BRST symmetry). Recall that the quadratic action had a zero-mode associated with proper time translation invariance. The gauge-fixing factor introduced in equation \eqref{fp} can be represented as follows
\begin{equation} \label{gffactor}
	\frac{1}{\omega} \int_0^1 \! dt \, \delta\left( g(t) \right) \frac{d}{dt} g(t) = \frac{1}{\omega} \int\! dt \int [dB\,dc\, d\bar{c}]  ~ e^{-\left[ 2\pi i B\cdot g(t) + \bar{c} c \cdot \frac{d}{dt} g(t) \right]},
\end{equation}
where $c$ and $\bar{c}$ are constant, anti-commuting Faddeev-Popov ghosts, and $B$ is a Lagrange multiplier\footnote{
The gauge-fixed action $S_{gf}$, which now includes the exponent in \eqref{gffactor}, enjoys as usual a BRST symmetry
\begin{equation}
	\hat \delta \, S_{gf} = 0,
\end{equation}
where
\begin{equation}
	\hat\delta \, t = - c, \quad \hat\delta \,\bar{c} = 2\pi i B, \quad \hat\delta\, c = \hat\delta\, B = \hat\delta\, (\dT) = 0 .
\end{equation}
Recall that $t$ parametrizes the gauge transformation, $\d z_t (\sigma, \tau) \equiv \d z (\sigma, \tau+t)$. The gauge-fixing part introduced in \eqref{gffactor} can be written as the following BRST exact expression:
$$
\hat\delta \left[\bar{c} \,g(t) \right] .
$$
}. Thus after using the gauge-invariance of the path integral to translate $t$ to zero, we can write the full path integral as follows
\begin{equation}
	\Gamma = \frac{2}{\omega V} \Im ~ \sum_{k=1}^\infty \int [d(\d x_\mu)] d (\dT) dB dc d\bar c ~ \frac{\eta^2 \left(\frac{i/2}{T_0\k + \dT} \right)}{T_0\k + \dT} ~ e^{-S( X_0\k+\d X, T_0\k + \dT )
    - \left[ 2\pi i B\cdot g(0) + \bar{c} c \cdot \dot g(0) \right]}	
\end{equation}

In order to construct the sought-after transformation, let us specify our gauge-fixing function. For present purposes, a judicious choice will be
\begin{equation}
	g(t) = \frac{2\pi k}{\kappa} \left[ \ev{z_t|\dot z_0} - \frac{1}{(2\pi k)^2} \frac{1}{T_0 T} \ev{z_t' |\dot z_0'} \right]
\end{equation}
where $z_t \equiv z(\sigma,\tau+ t)$ and $\kappa$ is defined as
\begin{equation} \label{kappa}
	\kappa \equiv \frac{1}{\modd{z_0}} \left( \modd{\dot z_0}^2 - \frac{1}{T_0^2} \modd{z_0'}^2 \right) = \frac{(2\pi k R)^2}{\modd{z_0}}, \qquad R\equiv d/\pe .
\end{equation}
This differs from our earlier choice \eqref{gaugefixfn} by the addition of the second term, but it is normalized such that the leading order Jacobian $J_{FP}^{\text{classical}}$ is the same as before. The Gribov factor is still $\omega = 2k$. We will argue that the fluctuation part can ultimately be set to zero. The constraint function becomes
\begin{eqnarray} \label{gofzero}
	g(0) &=& \frac{2\pi k}{\kappa} \left[ \ev{\d z | \dot z_0} - \frac{1}{(2\pi k)^2} \frac{\ev{ \d z' |\dot z_0'}}{T_0 T} \right] \nonumber \\
    &=& \frac{2\pi k}{\kappa} \left[ \frac{\z}{\modd{\dot z_0}} \left( \modd{\dot z_0}^2 - \frac{\modd{z_0'}^2}{T_0 T} \right) \right] - f(\dT, \{\y_i\}) \nonumber \\
    &=& \z \left[ 1 - \frac{\kappa}{4\pi\alpha'} \dT \, h(\dT) \right] - f(\dT, \{\y_i\}) \label{g-of-zero}
\end{eqnarray}
where the function $f$ depends on all modes except $\v$ and $\z$, while $h$ depends only on $\dT$. Our symmetry argument will eliminate the second term in square brackets, so that the constraint will reduce to $\delta (\z-f)$ with unit coefficient in front of $\z$. Similary, the Faddeev-Popov jacobian becomes
\begin{subequations} \label{jfp}
\begin{eqnarray}
	J_{FP}(z,T) \equiv \dot g(0) &=& \frac{2\pi k}{\kappa} \left[ \ev{\dot z | \dot z_0} - \frac{\ev{z' | z_0'}}{T_0 T} \right] \\
    &=& \modd{\dot z_0} + \ldots
\end{eqnarray}
\end{subequations}
Expanding the full gauge-fixed action $S_{gf}$ about the $k$'th instanton, we have
\begin{equation}
	S_{gf}\k = S_{\text{classical}}\k + S_{\text{quad}}\k + S_{\text{int}}\k ,
\end{equation}
$S_{\text{classical}}\k$ is given by \eqref{Sclassical}, and using \eqref{Sfluc},\eqref{gffactor},\eqref{gofzero} and \eqref{jfp}, we have\footnote{
Recall that uppercase $X_\mu$ stands for all spacetime components, $\mu=0,\ldots,25$. Lowercase $x_\mu$ has $\mu=2,\ldots,25$, i.e. 
$$\modd{\d \dot X}^2 = \modd{\d \dot z}^2 + \modd {\d \dot x}^2$$
}
\begin{eqnarray}
	S_{\text{quad}}\k &=& \frac{1}{4\pi\alpha'} \left[ T_0 \modd{\d \dot X}^2 + \frac{1}{T_0}\modd{\d X'}^2 \right] +\frac{\dT}{2\pi\alpha'} \left[ \ev{\dot z_0|\d \dot z} -\frac{1}{T_0^2} \ev{z_0'|\d z'} \right] 
+ \left. iE\int_0^1 \! d\tau \, \delta\bar z\delta\dot z\right|_{\sigma=0}
\nonumber \\
	&&  +\frac{1}{2} a \dT^2 + 2\pi i B (\z - f) + \bar c c J_{FP}^{\text{classical}} \nonumber\\ 
	S_{\text{int}}\k &=& \frac{\dT}{4\pi\alpha'} \left[ \modd{\d \dot X}^2 - \frac{1}{T_0^2} \modd{\d X'}^2 \right] -  \frac{1}{4\pi\alpha'} \sum_{j=2}^\infty \frac{(-\dT)^j}{T_0^{j+1}} \Big[ 2\ev{X_0'|\d X'} + \modd{\d X'}^2 \Big]  \\
    &&- \frac{\kappa}{4\pi\alpha'} 2\pi i B \z \dT \cdot h(\dT) + \bar c c\, \big[J_{FP}(z,T) - \modd{\dot z_0} \big]
\end{eqnarray}
(Although $f$ is not quadratic, we include it in $S_\text{quad}$ for convenience since the final result will be independent of $f$). Similarly, we expand the measure around $T=T_0\k$:
\begin{equation}
	\frac{1}{T} \eta^2 \left[ i/2T \right] = \frac{1}{T_0\k} \eta^2 \left[ i/2T_0\k \right] e^{ - F\k(\dT)} = \frac{1}{T_0\k} \eta^2 \left[ i/2T_0\k \right] \left( 1 + \mathcal{O} (\dT) \right)
\end{equation}
where the fluctuation factor $F\k$, given by
\begin{eqnarray}
	F\k(\dT) &=& \log \left[ \frac{\eta^2(i/2T)}{(1+\dT/T_0\k) \eta^2(i/2T_0\k)} \right] \\
    &=& \sum_{j=1}^\infty F_j\k \dT^j ,
\end{eqnarray}
contains all corrections to the measure, including the reparametrization ghost contribution. The Taylor expansion starts at $\dT^1$, and is well-defined since $\eta(x)$ is a holomorphic function in the upper half-plane.
The amplitude is therefore 
\begin{equation}
	\Gamma(\lambda) \equiv \frac{2}{\omega V} \Im ~ \sum_{k=1}^\infty e^{-S_{\text{classical}}\k} \int_{-\infty}^\infty \! d (\dT) \int [d(\d x_\mu)] dB dc d\bar c~ \frac{\eta^2 \left[ i/2T_0\k \right]}{T_0\k} ~ e^{-S_{\text{quad}}\k - \lambda \left(  S_{\text{int}}\k + F\k(\dT) \right)}
\end{equation}
evaluated at $\lambda = 1$. Moreover, the semiclassical approximation to $\Gamma$ that we computed in section \ref{instanton} is given by $\Gamma(0)$.

In what follows, we will demonstrate that $\frac{\partial}{\partial\lambda} \Gamma(\lambda)=0$ and therefore $\lambda$ can be deformed to zero without altering the value of $\Gamma$. From now on we drop the superscript $(k)$.

Define the nilpotent fermionic transformation $\Delta$ by
\begin{equation}
	\Delta \bar c = \frac{\kappa}{4\pi\alpha'} \dT, \qquad \Delta \, \d z (\sigma,\tau) = - \frac{1}{2} (2\pi k)  \, z_0(\sigma,\tau)\cdot c, \qquad \Delta(\text{other}) = 0.
\end{equation}
with $\kappa$ given by \eqref{kappa}. It leaves the quadratic action invariant:
\begin{equation}
	\Delta S_{\text{quad}}\k = 0 .
\end{equation}
We will show that $S_{\text{int}}$ is $\Delta$-exact (and therefore in particular it is $\Delta$-closed: $\Delta S_{\text{int}} =0$). In a given ($k$-)instanton sector, all bosonic interaction terms are contained in the combination
\begin{eqnarray}
	\mathcal {T} &\equiv & S(T,X) - S(T_0\k,X) + F\k(\dT) \\
    &=& \frac{\kappa}{4\pi\alpha'} \dT \cdot \xi\,[T, X]  
\end{eqnarray}
where we have defined
\begin{equation} 
	\xi \equiv \frac{1}{\kappa} \left(\modd{\dot X}^2 -\frac{1}{T_0 T} \modd{X'}^2 \right) + \frac{4\pi\alpha'}{\kappa} F\k(\dT) \label{xi}
\end{equation}
We showed earlier that in the absence of interaction terms, the path integral does not depend on the $\dT^2$ term or the $\dT\cdot\y_i$ cross terms. Therefore the semi-classical approximation corresponds, in the bosonic sector, to replacing $\xi \rightarrow 2\v$.

It is now easy to see that $\mathcal T$ and $\bar c c \dot g(0)$ are generated by $\Delta$ as follows:
\begin{equation} \label{deltacxi}
	\Delta \left( \bar c \, \xi \right) = \mathcal T + \bar c c \, J_{FP} ,
\end{equation}
while the correction to the ``constraint'' term can be written
\begin{equation}
	\Delta \left( - \bar c \cdot  2\pi i B \z h(\dT) \right) = - \frac{\kappa}{4\pi \alpha'} 2\pi i B \z \dT \cdot h(\dT)
\end{equation}
We have thus written a part of the action, including $S_\text{int}$, as a $\Delta$-exact quantity. However, we cannot simply set this to zero, as all $\v$ and $\bar c c$ dependence would be eliminated, i.e. we would get $\Gamma\sim 0\cdot \infty$. Before localizing we must separate out the quadratic part $(\bar c c J_{FP}^{\text{classical}} + \frac{\kappa}{4\pi\alpha'}\dT \v)$ of \eqref{deltacxi} by substracting off its preimage under $\Delta$ on the left hand side, namely
\begin{equation} \label{xi0}
	\xi_0 \equiv \frac{2}{\kappa } \left( \ev{\d \dot X |\dot X\cl} - \frac{1 }{T_0^2}\ev{\d X'| X\cl} \right); \qquad 
    \Delta \xi_0 = \bar c c J_{FP}^{\text{classical}} + \frac{\kappa}{4\pi\alpha'}\dT \v.
\end{equation}
Finally, we have
\begin{eqnarray}
	S_{\text{int}}(\dT,\d X) + F\k(\dT) &=& \left(\mathcal T - \frac{\kappa}{4\pi\alpha'} \dT \xi_0 \right) + \bar c c \left( J_{FP}-J_{FP}\cl \right)	+ 2\pi i B \z \frac{\kappa}{4\pi\alpha'} \dT \cdot h(\dT) \nonumber \\
    &=& \Delta \psi
\end{eqnarray}
where
\begin{equation}
	\psi \equiv  \bar c \left(\xi - \xi_0 - 2\pi i B \z h(\dT) \right)
\end{equation}
and $\xi$, $\xi_0$ and $h(\dT)$ were defined in \eqref{xi}, \eqref{xi0} and \eqref{g-of-zero} respectively.
Consequently,
\begin{eqnarray}
	\frac{\partial}{\partial \lambda} \Gamma(\lambda) &=&
    \frac{2}{\omega V} \Im ~ \sum_{k=1}^\infty \int \! [d(\d X_\mu)] d (\dT) dB dc d\bar c~ \frac{\eta^2 \left[ i/2T_0\k \right]}{T_0\k} (\Delta \psi) e^{-S_{\text{classical}}\k - S_{\text{quad}}\k - \lambda \Delta \psi} \nonumber \\
    &=& \frac{2}{\omega V} \Im ~ \sum_{k=1}^\infty \int \! [d(\d X_\mu)] d (\dT) dB dc d\bar c ~ \Delta \left\{  \frac{\eta^2 \left[ i/2T_0\k \right]}{T_0\k} \, \psi \, e^{-S_{\text{classical}}\k - S_{\text{quad}}\k - \lambda \Delta \psi} \right\} \nonumber \\
    &=& 0 .
\end{eqnarray}
We have thus proven that the full, interacting path integral \eqref{cylinderamp} is given exactly by its semi-classical approximation.

\section{Discussion}

In this paper we have studied the open bosonic string where the D-brane on which the string ends contains a constant electric field. We have examined the amplitude for string pair creation by tunnelling, the analog of the Schwinger effect for charged particle-antiparticle pairs in an electric field
. The string theory tunnelling process is mediated by instantons of the string sigma model which computes the open string annulus amplitude.   We have analyzed fluctuations about the classical  multi-instanton solutions and we found that integrating the Gaussian fluctuations and summing over all possible multi-instanton configurations obtains the known formula for the amplitude.   This can be regarded as another confirmation of that formula.  
We call the approximation which retains the classical instanton action and the determinants due to the quadratic fluctuations the WKB limit and our first result suggests that the WKB limit is exact. We have then fashioned a localization argument to prove that it is indeed exact.   The cohomology used for the localization of the functional integral utilizes the Fadeev-Popov ghosts which arise from the introduction of a particular collective coordinate, but it differs in form from the usual BRST cohomology.

The fermionic symmetry which we find could be useful in computing correlation functions, for example, even in integrals which are not WKB exact, it could
be used for re-organizing the integral to a more convenient form. We leave a detailed study of whether this can indeed be used to simplify perturbative computations of more complicated scenarios such as those with non-constant electric fields, to future study.
The problem of string pair production in non-homogeneous background fields has, to our knowledge, only been addressed in a few works to date \cite{Durin2003,Bolognesi2016,Condeescu2017} and demands further exploration.

Finally let us point out some appealing connections  with the literature.  Under a T-duality transformation in the $\vec E$ direction, our D$p$-brane setup is mapped to a pair of D$(p-1)$-branes with constant relative velocity (but zero extension) in the $\vec E$ direction.  This establishes a connection to questions of D-brane dynamics, see for example \cite{Bachas1996,Bachlechner2013,DAmico2015}. From this perspective, pair production occurs due to time dependence of the brane separation and, consequently, the open string spectrum. The critical electric field \eqref{critical_electric_field} manifests itself as a limiting velocity for the relativistic mechanics of D-branes, namely the speed of light.

Our work here may serve as a useful toy model for related calculations in curved space, in particular in the holographic context, for example involving the instanton fluctuation prefactor in the holographic Schwinger effect \cite{Semenoff2011}, and perhaps even the meson decay process analyzed in \cite{Faulkner2009}\footnote{
In the latter, mesons at finite temperature are modeled by strings connected to a D7 ``flavor'' brane outside of a black hole in $AdS_5\times S^5$. Dissociation of the mesons is mediated by worldsheet instantons, which allow for leaking of mesons into the black hole.
}. Another intriguing application is to pomeron physics. In \cite{Basar2012a}, high energy, inelastic scattering of dipoles in holographic QCD is studied and is found to be well modeled in a certain ``soft pomeron'' regime by D0-brane scattering in flat space. This is in turn related -- via the T-duality just discussed -- to Schwinger pair production mediated by worldsheet instantons. 

Lastly, it could be interesting to consider a hybrid particle-string version of the model \eqref{stringactiona}, wherein one includes a particle-like kinetic term on the boundary in addition to the gauge field coupling and string bulk terms. On the one hand, such a model provides a kind of interpolation between the string and particle models considered in this chapter and the previous one (see appendix A of \cite{Tseytlin1999} where such a calculation is performed explicitly for the string disk worldsheet). On the other hand, it was shown in \cite{Callan1990} by integrating out the free bulk degrees of freedom of the worldsheet that such a setup is equivalent to the dissipative quantum mechanics of Caldeira and Leggett \cite{Caldeira1983,Caldeira1983a}. This permits one to study, for example, pair nucleation of particles in a dissipative setting \cite{Acatrinei1999}.

\vskip .25cm
\noindent
The authors acknowledge the financial support of NSERC. 
 
\appendix

\section{A toy model} \label{toymodel}

Let us consider the simple model with an ordinary integral where we want to compute the imaginary part.   
Consider the integral
\begin{align}
Z=\Im \int_0^\infty \frac{1}{T} \int d^2z e^{-(T-T_0)\bar z z -M^2/T}
\label{z}
\end{align}
The integral measure and the integrand are real and positive, so the reader might wonder where it gets 
an imaginary part.   To understand this, notice that, when $T_0$ is real and positive, the  
coefficient of the quadratic form $\bar z z$ in the integrand is not positive for all values of $T$, if $T<T_0$ it
is negative and the $z$-integration would diverge.  To define the integral, we first assume that $T_0$ is negative,
perform the integration which is then well-defined, and obtain a function of $T_0$ which can be analytically
continued to the complex $T_0$ plane.  It is this analytic continuation which produces an imaginary part for
the integral when $T_0$ is positive.  

Explicitly, we first do the Gaussian integral over $z$ to get
\begin{align}
Z=\Im \int_0^\infty \frac{1}{T} \frac{\pi}{T-T_0-i\varepsilon} e^{ -M^2/T} =  \frac{\pi^2}{T_0}e^{ -M^2/T_0} 
\label{z1}
\end{align}
Then, if $T_0$ is real and positive, we define the  distribution in the integrand by taking $T_0$ to the real
axis from the upper half-plane   by replacing $ \frac{\pi}{T-T_0} $ with
$ \frac{\pi}{T-T_0-i\varepsilon} $.  We can then use the identity $\Im  \frac{1}{T-T_0-i\varepsilon} = \pi\delta(T-T_0)$.  
This allows us to find the exact imaginary part
\begin{align}
Z =  \frac{\pi^2}{T_0}e^{ -M^2/T_0} 
\label{z2}
\end{align}
Alternatively, we could consider an approximate evaluation of the integral by saddle point technique.  Such an approximation
should be accurate when the ``action''
\begin{align}
S=(T-T_0)\bar z z +M^2/T 
\end{align}
is large, that is then $\frac{M^2}{T_0}>>1$.   To implement the saddle-point technique, we consider the classical
equations of motion, 
\begin{align}
\bar z z - M^2/T^2=0 ~~,~~(T-T_0)z=0
\label{cem}
\end{align}
and we find the classical solutions
$z= \frac{M}{T_0}$, $T=T_0$ and we expand 
the integration variables as the classical solutions plus
``fluctuations'',
\begin{align}
z= \frac{M}{T}+\delta z~~,~~T=T_0+\delta T
\label{fluc}
\end{align}
Before we proceed, we notice that the action has a flat direction, it is invariant under the phase
transformation $z\to e^{-i\theta}z$, $\bar z\to e^{i\theta}\bar z$.   Such a symmetry will lead to a zero mode
in the fluctuations about the classical solution.  We must take care of this symmetry,
and degeneracy of the solution by gauge fixing.  Most convenient is the Fadeev-Popov trick of 
inserting unity into the integration measure in equation (\ref{z}) in the following form, 
\begin{align}
1 = \frac{1}{2}\int_0^{2\pi} d\theta \delta\left( \frac{1}{2i}(e^{i\theta}\bar  z - e^{-i\theta}z)\right)\left|  \frac{d}{d\theta}
\left( \frac{1}{2i}(e^{i\theta}\bar  z - e^{-i\theta}z)\right)\right|
\label{fptoy}
\end{align}
The $\frac{1}{2}$ which appears in front of the right-hand-side arises from a Gribov copy -- there are two solutions of
the equation $e^{i\theta}\bar  z - e^{-i\theta}z=0$ in the interval $\theta\in[0,2\pi)$. One then removes $\theta$ by a symmetry
transformation.
The upshot is the  insertion of the following into the integral in equation (\ref{z}),
$$
\pi \delta(\Im z) |\Re z|=  \int  d\bar cdc db~ e^{-\pi \Re  z\bar c c-2\pi ib\Im z} 
$$
where $c$ and $\bar c$ are Fadeev-Popov ghosts.  It now takes the form
\begin{align}
Z=\Im  \int [dT d^2z d\bar c dc db] ~\frac{1}{T} e^{-(T-T_0)\bar z z -M^2/T-\pi\Re  z\bar c c-2\pi ib\Im z}
\label{zgf}
\end{align}
Now, we expand the action about the classical solution (with the classical part of the ghosts vanishing). It becomes
\begin{align}
&S=S_{0}+S_{\rm quad }+S_{\rm int } \nonumber \\
&S_0=\frac{M^2}{T_0} \label{gfactionexpanded1}\\
&S_{\rm quad }=  \frac{M}{T_0}\delta T( \delta z+\delta\bar z )+\frac{M^2}{T_0^3}\delta T^2 
+\frac{M}{T_0}\pi \bar c c   +2\pi i\Im z b    \label{gfactionexpanded2} \\
 &S_{\rm int }=  \delta T\delta\bar z\delta z
 +\pi\Re \delta  z\bar c c   
+\sum_{k=3}^\infty \frac{M^2}{T_0^{k+1}}(-\delta T)^k
\label{gfactionexpanded3}
\end{align}
The classical part of the action $S_{0}=\frac{M^2}{T_0}$ matches the exponent in   
  equation  (\ref{z2}). The corrections to this classical limit begin with dropping all terms of order higher than quadratic
in the fluctuations, that is, dropping $S_{\rm int }$,  and doing the remaining Gaussian integral over the fluctuations.  
The resulting integral is complex, due to the fact that the determinant of bosonic quadratic from is 
negative 
\begin{align}\label{bosonic_quadratic_form}
\det \left[ \begin{matrix}  0 &   \frac{M}{T_0} \cr \frac{M}{T_0} & 2\frac{M^2}{T_0^3} \cr \end{matrix} \right]= - \left(\frac{M}{T_0}\right)^2
\end{align}
and the gaussian integration contributes the inverse of the square root of this determinant.  This 
produces a factor of $i\pi\frac{T_0}{M}$ in the measure, 
 where we have chosen an appropriate sign for $i=\sqrt{-1}$.  
 The fermionic quadratic form contributes $\pi \frac{M}{T_0}$
so the net factor from integrating the fluctuations is $i\pi^2$. 
The remainder of the integrand is evaluated at the classical solution. 
The result of the WKB approximation to the integral in this ``instanton'' sector  is then
purely imaginary
$$
Z=\Im \left\{i\frac{\pi^2}{T_0}e^{ -M^2/T_0}   \right\}
$$
and it appears to get the imaginary part of the integral, compare with  (\ref{z2}), exactly.  
However, this should  be an approximation. There are
higher order than quadratic terms in the action,$S_{\rm int}$  as well as the factor $\frac{1}{T}$  in the measure,  
and there should be  corrections to this result.  Apparently,
if the two computations are to match,  such corrections  must cancel.  

To see how this can happen, we observe that the action in equations (\ref{gfactionexpanded1})-(\ref{gfactionexpanded3})
has a fermionic symmetry under the transformations
\begin{align}
	&\Delta \bar{c}  =-\delta T ~,~
	\Delta  \delta z =\Delta \delta\bar z  = \frac{\pi}{2}c ~,~\\
	&\Delta c=\Delta \delta T=\Delta b=0
\end{align}
This transformation is nilpotent , $\Delta^2({\rm anything})=0$ and also
$$
\int d\bar c c \Delta\left( {\rm ~anything ~even~ in ~}c,~\bar c\right)=0
$$
since the integral of any quantity which is odd in the Grassman numbers must vanish.

It is easy to check that $\Delta S=0$.  As well,  the 
measure factor, $\frac{1}{T}=\frac{1}{T_0}\frac{1}{1+\delta T/T_0}$,  is invariant 
$$
\Delta \left[\frac{1}{T_0}\frac{1}{1+\delta T/T_0}\right]~=~0
$$

Moreover, the ``interaction'' terms in the action can be written as transformations of simple functions, 
\begin{align}
 S_{\rm int }= \delta T \delta \bar z\delta z +\pi\delta  z\bar c c
+\sum_{k=3}^\infty \frac{M^2}{T_0^{k+1}}(-\delta T)^k 
=\Delta\left( -\delta\bar z\delta z\bar c     -\sum_{k=3}^\infty \frac{M^2}{T_0^{k+1}}(-\delta T)^{k-1}\bar c \right)\\
\label{interaction_action}
\end{align}
In addition the measure factor
$$
 \frac{1}{T_0}\frac{1}{1+ \frac{\delta T}{T_0}}= \frac{1}{T_0}+\Delta\left[ \sum_{k=1}^\infty \bar c \frac{(-\delta T)^{k-1}}{T_0^k}\right]
 $$
Then, 
\begin{align}
Z&=\int [dTd^2z d\bar cdcdb] ~ \frac{1}{T_0}\frac{1}{1+ \frac{\delta T}{T_0}}~e^{-S_0-S_{\rm quad }- S_{\rm int }}
 \nonumber\\&=\int[dTd^2zd\bar cdcdb]\left[ \frac{1}{T_0} ~e^{-S_0-S_{\rm quad } }+\Delta\left\{...\right\}\right]
 =\int[dTd^2zd\bar cdcdb]~ \frac{1}{T_0} ~e^{-S_0-S_{\rm quad } } 
\end{align} 
which tells us that the saddle point  approximation of the   integral is equal to the exact result in the ``instanton sector''.

\section{Gaussian integral computation} \label{gaussianintegral}
We can evaluate the partition function with action (\ref{stringaction}) and open string boundary conditions in two different ways. In section \ref{instanton} we considered the semi-classical limit and integrated both $T$ and the coordinates using a saddlepoint technique. In this appendix we shall do the Gaussian functional integral over the string embedding coordinates $(z,x_a,x_A)$ and then the integral over the modular parameter of the cylinder, $T$.  By picking up the residues at an infinite sequence of poles we shall reproduce the formula \eqref{finalresult} for the imaginary part of the vacuum energy \cite{Bachas1992}.

For this computation, we introduce the mode expansion
\begin{align} \label{modeexpansion}
&z(\sigma,\tau)= \sum_{n=-\infty}^\infty\sum_{m=1}^\infty  e^{2\pi in\tau}\sqrt{2}\cos m\pi\sigma ~z_{nm}  
+\sum_{n=-\infty}^\infty  e^{2\pi in\tau} ~z_{n0} \\
&x_a(\sigma,\tau)= \sum_{n=-\infty}^\infty\sum_{m=1}^\infty  e^{2\pi in\tau}\sqrt{2}\cos m\pi\sigma~x_{m n a} 
+ \sum_{n=-\infty}^\infty e^{2\pi in\tau} ~x_{an0}  \\
&x_A(\sigma,\tau)=\sigma\vec{d} +\sum_{n=-\infty}^\infty\sum_{m=1}^\infty  e^{2\pi in\tau}\sqrt{2}\sin m\pi\sigma ~x_{Anm}  
\end{align}
where we have used Neumann boundary conditions for $x_a$ and Dirichlet boundary conditions for $x_A$.  We will also use Neumann boundary conditions for the normal modes in which we expand $z(\sigma,\tau)$.  These are not the correct boundary conditions
for $z$. The result of using these boundary conditions is that the quadratic form which we will obtain is not diagonal.

When we plug the mode expansions into the the action, it becomes
\begin{align}
S&=\frac{\vec{d}^2}{4\pi\alpha'T} \nonumber\\
&+\sum_{m=0}^\infty\sum_{n=-\infty}^\infty\frac{\left[(2\pi n)^2T^2+(\pi m)^2\right]}{4\pi\alpha'T}|x_{anm}|^2  
+\sum_{m=1}^\infty\sum_{n=-\infty}^\infty\frac{\left[(2\pi n)^2T^2+(\pi m)^2\right]}{4\pi\alpha'T}|x_{Anm}|^2 \nonumber \\
&+\sum_{m=0}^\infty\sum_{n=-\infty}^\infty\frac{\left[(2\pi n)^2T^2+(\pi m)^2\right]}{2\pi\alpha'T}|z_{nm}|^2
-\sum_{n=-\infty}^\infty (2\pi nE) \, \left|z_{n0}+\sum_{m=1}^\infty \sqrt{2}z_{nm} \right|^2
\label{modeaction}
\end{align}
The path integral measure is
\begin{align}\label{stringmeasure}
\int_0^\infty \frac{dT}{2T}\int \left[ dX^\mu(\sigma,\tau) \right]
~[{\rm ghost}]~~=~V\int_0^\infty \frac{dT}{2T}\prod_{n=-\infty}^\infty\prod_{m=1}^\infty dz_{nm}d\bar z_{nm}dx_{anm}dx_{Anm}
\cdot \nonumber \\
\cdot
\prod_{n=-\infty,n\neq 0}^\infty dz_{n0}d\bar z_{n0}dx_{an0} 
 ~{\det}' \left[ -\frac{1}{T}\partial_\sigma^2-T\partial_\tau^2
\right]
\end{align}
where we have included the ghost determinant and the prime on the ghost determinant indicates that 
simultaneous zero mode 
of $\partial_\tau$ and $\partial_\sigma$ is omitted. 
The factor of the D-brane worldvolume $V$ is from
the integral over similar zero modes of $x_a$ and $z$, namely $x_{a00}$ and $z_{00}$.  

The  quadratic form in the last line of the action in equation (\ref{modeaction}) is a non-diagonal matrix.
This is not surprising, as we have
expanded in modes which do not obey the correct boundary condition.  To proceed, we will have to find the eigenvalues
of that matrix.  We can make this slightly easier to deal with by rescaling the modes by
\begin{align} \label{zscale}
z_{nm}\to \sqrt{\frac{ 2\pi\alpha' T}  {  (2\pi n)^2T^2+(\pi m)^2 } } \, z_{nm}~,\qquad
\bar z_{nm}\to \sqrt{ \frac{ 2\pi\alpha' T }   {(2\pi n)^2T^2+(\pi m)^2  } } \, \bar z_{nm}
\end{align}
The resulting action is
\begin{align}
	S&=\frac{\vec d^2}{4\pi\alpha'T} +\sum_{m=0}^\infty\sum_{n\neq 0}\bar z_{nm}\left[\delta_{mm'}-\E_{mm'}\right]z_{m'n} + \sum_{m=1}^\infty |z_{m0}|^2
\nonumber \\
&+\sum_{m=0}^\infty\sum_{n=-\infty}^\infty\frac{\left[(2\pi n)^2T^2+(\pi m)^2\right]}{4\pi\alpha'T}|x_{anm}|^2 
+\sum_{m=1}^\infty\sum_{n=-\infty}^\infty\frac{\left[(2\pi n)^2T^2+(\pi m)^2\right]}{4\pi\alpha'T}|x_{Anm}|^2 
\label{modeaction1}
\end{align}
where we define the matrix
\begin{align}\label{scripte}
\left[ \begin{matrix}
\E_{00}(n)& \E_{0m'}(n) \cr
\E_{m0}(n) &  \E_{mm'}(n)   \cr
\end{matrix}\right]   
&=2\pi n E\left[ \begin{matrix}
g(n,0) & \sqrt{2g(n,0)g(n,m')}  \cr
 \sqrt{2 g(n,m)g(n,0)} &  2\sqrt{ g(n,m)g(n,m')}   \cr
\end{matrix}\right] , \\
\label{g}
 g(n,m)&=\frac{2\pi\alpha'T}{(2\pi n)^2T^2+(\pi m)^2} \, .
\end{align}
Taking into account the Jacobian in the measure resulting from this rescaling, and then doing  the Gaussian integrals 
over the coordinates will result in the appearance of the determinants in the integrand:
\begin{align}
&Z=V \int_0^\infty\frac{dT}{2T}e^{-\frac{\vec d^2}{4\pi \alpha' T}}\prod_{n=-\infty}^\infty\prod_{m=1}^\infty(2\pi)^{12}\left[  \frac{\left[(2\pi n)^2T^2+(\pi m)^2\right]}{2\pi\alpha'T}\right]^{- {12}}
\cdot \nonumber \\ 
&\cdot 
\prod_{n\neq 0} (2\pi)^{ {\frac{D-1}{2}}}\left[\frac{(2\pi n)^2T}{2\pi\alpha'}\right]^{- {\frac{D+1}{2}}}~
\cdot 
\frac{1}{2\pi} \prod_{n\neq 0} \left[  \det\left[
\delta_{mm'}-\E_{mm'} \right]\right]^{-1}
\label{determinants}\end{align}
where the infinite products
$$
\prod_{n=-\infty}^\infty\prod_{m=1}^\infty(2\pi)^{12}\left[  \frac{\left[(2\pi n)^2T^2+(\pi m)^2\right]}{2\pi\alpha'T}\right]^{- {12}}
$$
which appear in the first line are the determinants arising from integrating  $x_{anm}$, a factor from scaling $z_{nm}, \bar z_{nm}$ and 
the ghost determinant. 
The determinant in the second line is from the integral over $z_{nm}$ and $\bar z_{nm}$. 
When infinite products diverge, we use $\zeta$-function regularization to define them (see appendix \ref{zeta}.  For example, in finding the determinant in the second line of (\ref{determinants}), we encounter a product over all of the modes of the factor $2\pi$, 
\begin{multline} \label{twopiprod}
\prod_{mn}(2\pi)=(\prod_{m\geq 1}2\pi)(\prod_{n\neq 0} \prod_{m\geq 0} 2\pi) 
= (2\pi)^{\zeta(0)} (\prod_{n\neq 0} (2\pi)\cdot (2\pi)^{\zeta(0)} )
= \frac{1}{\sqrt{2\pi}} (\prod_{n\neq 0} \sqrt{2\pi}) \\
=\frac{1}{\sqrt{2\pi}} (\prod_{n\geq 1} 2\pi)
=\frac{1}{\sqrt{2\pi}}  (2\pi)^{\zeta(0)}
=\frac{1}{2\pi} , \qquad
\end{multline}
which is the factor in front of the determinant.  

The infinite products in (\ref{determinants}) can be put in a more convenient form.  Using the results of appendix \ref{zeta} we have that the infinite product in the first line of (\ref{determinants}) reduces to the usual modular form
\begin{align}
\prod_{m=1}^\infty\prod_{n=-\infty}^\infty\left[\frac{\left[(2\pi n)^2T^2+(\pi m)^2\right]}{2\pi\alpha'T}\right]^{-12}&
 =\left[e^{-\pi/12T}\prod_{m=1}^\infty [1-e^{-\pi m/T}]^2\right]^{-12}
= \eta^{-24}(i/2T) &
\end{align}
where $\eta(\tau)$ is the Dedekind eta-function (see Appendix A). In addition, using zeta-function regularization, 
$$
\prod_{n=-\infty,n\neq 0}^\infty(2\pi)^{ {\frac{D-1}{2}}}\left[\frac{(2\pi n)^2T}{2\pi\alpha'}\right]^{- {\frac{D+1}{2}}}
=2\pi \prod_{n=1}^\infty\left[\frac{(2\pi n)^2T}{4\pi^2\alpha'}\right]^{-  (D+1 )}
=2\pi \left[\frac{T}{4\pi^2\alpha'}\right]^{{\frac{D+1}{2}}} .
$$
We are left with the last determinant in the second line of (\ref{determinants}). Observe that the matrix $\E$ defined in \eqref{scripte} is the outer product of a vector and its transpose:
\begin{equation}
	\E = \tanh(  \pe )  \coth (2\pi n T) v_mv_{m'} ,
\end{equation}
where the normalized vector $\vec v$ is given by
\begin{equation} \label{vmode}
	v_m = \sqrt{\frac{n}{\alpha^\prime} \tanh(2\pi nT ) } (\sqrt{2})^{1-\delta_{m0}} \sqrt{g(n,m)} .
\end{equation}
It therefore has only one non-zero eigenvalue, namely 
$$
2\pi \alpha^\prime E \coth (2\pi n T) = \tanh(  \pe)  \coth (2\pi n T)
$$ 
Inserting this result into the partition function, and after transforming the integration variable $T\to 1/2T$, we find the expression
\begin{align}
 Z=V\int_0^{\infty} \frac{dT}{T}\frac{ e^{-T\vec d^2/2\pi\alpha'} \eta^{-24}(iT) \left[ 8\pi^2\alpha' T  \right]^{ -\frac{D+1}{2} }   }
{ \prod_{n=1}^\infty  
\left[ 1-( 2\pi\alpha'E)^2\coth^2 (\pi n /T) \right]     }
\end{align}
When $E\to 0$ this expression approaches the usual cylinder amplitude of the open bosonic string suspended between
two D-branes. Moreover, the integrand now has poles at discrete values of $T$, 
\begin{align}
T_k = \frac{k}{\varepsilon}~~(k\in\mathbb{Z}^+),~~\tanh \pi\varepsilon\equiv 2\pi\alpha' E
\end{align}

To proceed, it is necessary to find the residues of the poles.   For this purpose,  
it is convenient to perform a modular transformation of the
infinite product.   In order to do this, it is convenient to first write the infinite product  
as a product representation of Jacobi theta functions\footnote{See Appendix A for a definition of
the relevant theta function and its modular transformation property.} and then to use the known modular
transformation property of the theta function.  For this purpose, we use
the following sequence of manipulations,
$$
 \prod_{n=1}^\infty  \frac{1}{
\left[ 1-\tanh^2\pi\varepsilon\coth^2 (\pi n /T) \right] } =
 \prod_{n=1}^\infty  \frac{\cosh^2\pi\varepsilon\sinh^2(\pi n/T) }
{
\left[   \sinh (\pi n/T-\pi\varepsilon) \sinh (\pi n/T+\pi\varepsilon) \right] 
 }
$$
$$
 =\frac{1}{\cosh\pi\varepsilon}
\prod_{n=1}^\infty  \frac   
   {     [ 1-e^{-2\pi n/T} ]^3   } { [ 1-e^{-2\pi n/T} ]  [1-e^{-2(\pi n/T-\pi\varepsilon) } ]  [1-e^{-2(\pi n/T+\pi\varepsilon)} ] }
=\frac{2}{i \coth\pi\varepsilon}\frac{\eta^3(i/T) }{\Theta_{11}(i\varepsilon|i/T) }
$$
Using the modular transformation of the theta- and eta-functions, we find
\begin{align}
& \prod_{n=1}^\infty  \frac{1}{
\left[ 1-( 2\pi\alpha'E)^2\coth^2 (\pi n /T) \right]}   =2T\tanh\pi\varepsilon \cdot
 e^{-\pi\varepsilon^2T}\frac{\eta^3(iT) }{\Theta_{11}(-\varepsilon T | iT) }
\nonumber \\
&=\tanh\pi\varepsilon\frac{Te^{-\pi\varepsilon^2T}}{\sin(\pi\varepsilon T)}
\prod_{n=1}^\infty  \frac{[1-e^{-2\pi T n}]^2}{
(1-e^{-2T\pi (n+i\varepsilon) })(1-e^{-2 T \pi (n-i\varepsilon) })  }
\label{modularproduct}
\end{align}
Now, the factors in the infinite product are regular for nonzero real values of $T$.  
The poles on the real $T$-axis originate from the factor of the inverse of $\sin(\pi\varepsilon T)$ which
is outside of the infinite product.   The poles occur at $T= k/\varepsilon$ and,  at each pole, the residue is
$$
(-1)^k ke^{-\pi k\varepsilon }\frac{ \tanh\pi\varepsilon}{\pi\varepsilon^2}
$$ 
Note that the remaining infinite product is simply equal to one at the position of the pole. 
The imaginary part of the partition function is then given by a sum over (half-) residues at the poles, 
\begin{align}
 Z=\Re (Z) + V \frac{1}{2} 2\pi i \sum_{k=1}^\infty \frac{1}{2T_k} (-1)^k ke^{-\pi k\varepsilon }\frac{ \tanh\pi\varepsilon}{\pi\varepsilon^2} \frac{e^{-k\vec d^2/2\pi\alpha' \epsilon}}{\eta^{24}(ik/\epsilon)} \left[ 8\pi^2\alpha' k/\epsilon  \right]^{ -\frac{D+1}{2} }
\end{align}
so that the rate of pair production is
\begin{align}
	\Gamma = \frac{2}{V} \Im (Z) = \frac{ \tanh\pi\varepsilon}{\varepsilon}
\sum_{k=1}^\infty  \frac{   (-1)^{k+1}  e^{  -\pi k M^2 \frac{ 2\alpha'}{\varepsilon} -k\pi\varepsilon }      } {
 \eta^{24}(i k/\varepsilon)  }
\left[ \frac{\varepsilon/2\alpha'}{4\pi^2 k } \right]^{ \frac{D+1}{2} } .
\label{stringamp}
\end{align}

\section{Fluctuation prefactor from explicit mode expansion} \label{neumannmodeexpansion}
In section \ref{instanton} we evaluated the quadratic fluctuation prefactor using the Gelfand-Yaglom approach for functional determinants. In this appendix we present, for completeness, a ``brute-force'' calculation of the same fluctuation integral using an explicit mode expansion. Since the eigenvalues of the $\mu = 0,1$ fluctuation operator are determined by a transcendental equation, and therefore not known explicitly, we use modes for $\delta z$ which do not obey the correct boundary condition, but would be appropriate to the same
problem with Neumann boundary conditions. This yields a non-diagonal quadratic form. With a little work, and some cavalier manipulations of infinite matrices, we are able to find its determinant, as well as extract the zero- and tachyonic modes, to obtain finally the result \eqref{stringschwinger}.

Returning therefore to equation \eqref{Sfluc}, we now expand the fluctuations in modes as
\begin{align}
\delta z(\sigma,\tau) &= \sum_{n=-\infty}^\infty\sum_{m=1}^\infty e^{2\pi i n\tau}\sqrt{2}\cos \pi m\sigma \,\delta z_{nm}
+  \sum_{n=-\infty}^\infty e^{2\pi i n\tau}\,  \delta z_{n0}
\\
\delta x_a (\sigma,\tau)&= \sum_{n=-\infty}^\infty\sum_{m=1}^\infty e^{2\pi i n\tau}\sqrt{2}\cos \pi m\sigma \delta x_{anm}
+  \sum_{n=-\infty}^\infty e^{2\pi i n\tau} \delta x_{an0}
\\
\delta x_A (\sigma,\tau)&= \sum_{n=-\infty}^\infty\sum_{m=1}^\infty e^{2\pi i n\tau}\sqrt{2}\sin \pi m\sigma \delta x_{Anm}
\end{align}
Even though $\delta z$ has the wrong boundary conditions,  the linear terms in $\delta z$  vanish due to the fact that the classical solution obeys the equation of motion and it has the correct boundary conditions. Using this mode expansion and the equations of motion the action becomes
\begin{align}
S&= k m_0^2\frac{2\pi\alpha'}{\varepsilon}  + a \delta T^2+\frac{d}{2 \pi \alpha' T_0^2}  \sinh(\pe) \delta T \sum_{m=0}^{\infty} q_m \frac{(\delta z_{km}+\delta \bar z_{km})}{\sqrt{2}}\nonumber \\
&+\sum_{m=0}^\infty\sum_{n=-\infty}^\infty\frac{|\delta z_{nm}|^2}{2\pi\alpha'T_0}[T_0^2(2\pi n)^2+(\pi m)^2]
-\sum_{n=-\infty}^\infty 2\pi n E~\left|\delta z_{n0}+\sum_{m=1}^\infty\sqrt{2} \delta z_{nm}\right|^2 
 \nonumber \\
&+\sum_{m=0}^\infty\sum_{n=-\infty}^\infty\frac{|\delta x_{anm}|^2}{4\pi\alpha'T_0}[T_0^2(2\pi n)^2+(\pi m)^2] 
\nonumber \\
&+\sum_{m=1}^\infty\sum_{n=-\infty}^\infty\frac{|\delta x_{Anm}|^2}{4\pi\alpha'T_0}[T_0^2(2\pi n)^2+(\pi m)^2]
+\ldots
\end{align}
where
\begin{equation} \label{qm}
	q_m \equiv (\sqrt{2})^{1-\d_{m0}} \frac{(\pe)^2-\pi^2m^2}{(\pe)^2+\pi^2m^2}, \qquad a \equiv \frac{d^2}{2\pi\alpha'}\frac{1}{4T_0^3}\left(1
	+\frac{\cosh (\pe) \sinh (\pe) }{ \pe} \right)
\end{equation}

\subsection*{Transverse fluctuations and ghosts}

Performing the Gaussian integral over the coordinates  $\delta x_{a,A}$ yields the following factors in the path integral measure, 
for each of the modes with $m\neq 0$
\begin{equation}
	\prod_{n=-\infty}^{\infty} \prod_{m=1}^\infty \left[ \frac{2\pi \alpha' T_0}{(2\pi n)^2 T_0^2 + (\pi m)^2}\right]^{12} = \boxed{\eta^{-24}(i/2T_0) }
\end{equation}
where $\eta(z)$ is the Dedekind eta-function. 
For the modes with $m=0$
\begin{align}
	\prod_{n\neq 0} (2\pi)^\frac{D-1}{2} \left[ \frac{2\pi \alpha'}{(2\pi n)^2 T_0}\right]^\frac{D-1}{2} = \prod_{n>0} \left[\frac{4\pi^2 \alpha'}{(2\pi n)^2 T_0} \right]^{D-1} \nonumber \\= 
	 \left[\frac{4\pi^2 \alpha'}{(2\pi )^2 T_0} \right]^{(D-1)\zeta(0)}e^{2(D-1)\zeta'(0)} =
	\boxed{\left[\frac{T_0}{4\pi^2 \alpha'}\right]^\frac{D-1}{2} }.
\end{align}
where we have used zeta-function regularization to define the formally divergent infinite product (see appendix \ref{zeta}). 
Evaluating the ghost determinant (see discussion of reparametrization ghosts in section \ref{instanton}) yields the factor
\begin{equation}
	\det \left[-\frac{1}{T_0} \partial_\sigma^2 -T_0 \partial_\tau^2 \right] = \prod_{n=-\infty}^\infty \prod_{m=1}^{\infty} \frac{(2\pi n)^2 T_0^2 + (\pi m)^2}{2\pi \alpha' T_0} = \boxed{ \eta^{+2}(i/2T_0) }
\end{equation}

\subsection*{Lightcone-coordinate fluctuations}

Consider the quadratic form containing the variables $\delta z_{nm}$.  In order to evaluate its determinant, we shall
have to find its eigenvalues.   In order to define  its eigenvalues  it is convenient to rescale all $\d z_{nm}$ where either $m$ or $n$ is nonzero as
\begin{equation} \label{gscale}
\delta z_{nm}\to 
\sqrt{ g(n,m)}\,\delta z_{nm},	\qquad {\rm with} \quad  g(n,m)= \frac {2\pi\alpha'T_0}{T_0^2(2\pi n)^2+(\pi m)^2}
\end{equation}
This rescaling results in a Jacobian in the measure, (where the second product arises from the $m=0$ modes)
\begin{multline}
	\lefteqn{\prod_{n=-\infty}^\infty\prod_{m=1}^\infty \left[ \frac {2\pi\alpha'T_0}{T_0^2(2\pi n)^2+(\pi m)^2}\right]
\prod_{n=1}^\infty\left[ \frac {2\pi\alpha'T_0}{T_0^2(2\pi n)^2}\right]^2 } \\
	= \left\{\prod_{n=-\infty}^\infty\prod_{m=1}^\infty \left[ \frac {2\pi\alpha'T_0}{T_0^2(2\pi n)^2+(\pi m)^2}\right]\right\}
\left[ \frac {2\pi\alpha'T_0}{T_0^2(2\pi )^2}\right]^{2\zeta(0)}e^{4\zeta'(0)} \\
= \boxed{ \frac{2\pi}{\eta^2(i/2T_0)}\frac{T_0}{4\pi^2\alpha'} } \label{Jg}
\end{multline}
This leaves the following quadratic form in the action
\begin{align}
	S_{\rm quadratic}&= a \delta T^2 
+ \frac{d }{2 \pi \alpha' T_0^2}  \sinh(\pe ) \delta T \sum_{m=0}^{\infty} q_m \sqrt{g(k,m)} \frac{(\delta z_{km}+\delta \bar z_{km})}{\sqrt{2}}
\nonumber \\
&+\sum_{m=0}^\infty\sum_{n\neq0}    \delta\bar z_{nm}  \left[ \delta_{mm'}-  \E_{mm'} (n)\right]   \delta z_{nm'}  
+\sum_{m=1}^\infty\ |\delta z_{0m}|^2 
+\ldots \label{tzquadratic}
\end{align}
where
\begin{align} \label{Ematrix}
\left[\begin{matrix} \E_{00}(n) &\E_{0m'}(n)\cr \E_{m0}(n)& \E_{mm'}(n)\cr
 \end{matrix}\right]=&   2\pi n E \left[\begin{matrix}  g(n,0) & \sqrt{2g(n,0)g(n,m')} \cr\sqrt{2g(n,m)g(n,0)}
& 2\sqrt{g(n,m)g(n,m')}\cr\end{matrix}\right]~~,~~m,m'=0,1,2,...   \\
=& \tanh(  \pe)  \coth (\pe n/k) v_mv_{m'}
\end{align}
This is just the matrix in \eqref{scripte}, evaluated at $T=T_0$. Consequently the quadratic form $(\mathcal{I}-\E(n))$ has eigenvalues $1$ and $1-\tanh(\pe)  \coth (\pe n/k)$. When $n=k$ the latter is zero. The integration of the modes $\delta z_{nm}$ with $n=k$ excluded produces the factor 
\begin{eqnarray}
	\prod_{n\neq 0,k}\frac{1}{\det\left({\cal I}-\E(n)\right)}&=&\frac{1}{\prod_{n\neq 0,k}\left[ 1-\tanh(  \pe)  \coth (\pe n/k)\right] } \nonumber \\
&=&\boxed{\frac{(-1)^{k+1}e^{-\pi k\varepsilon}}{\cosh^2\pi\varepsilon }}
\end{eqnarray}
which is derived in appendix \ref{zeta}, equation \eqref{appendixmodproduct}.

\subsection*{Tachyon: real $n=k$ modes coupled to $\delta T$}
Now consider $n=k$, for which $(\mathcal{I}-\E)$ has a zero eigenvalue. Note that $\delta T$ couples only to the real part of $\delta z_k$. More precisely, we have that $\frac{\delta z_{km}+\delta \bar{z}_{km}}{\sqrt{2}}$ and $\delta T$ are coupled in the following quadratic form:
\begin{equation}
\mathcal{M} = \left(
\begin{array}{cc}
 a & \vec{J}^{\mathsf{T}} \\
 \phantom{m}\vec{J}\phantom{m} & \mathcal{I}-\E_k \\
\end{array}
\right)
\end{equation}
where the $\delta T$, $\delta z$ cross-term is 
\begin{equation}
	J_m=\frac{d \sinh (\pi \epsilon ) }{4\pi  \alpha'  T_0^2 } \, q_m \sqrt{
   g(k,m)} 
\end{equation}
and $q_m$ is given by \eqref{qm}. We found the spectrum of the submatrix $(\mathcal{I}-\E)$ above. In the gaussian integration over the quadratic form $\mathcal{M}$, we will find that although $\delta T$ couples to all eigenmodes of $(\mathcal{I}-\E)$, only its coupling to the would-be zero mode $\v$ contributes to the Gaussian integral. We demonstrate this as follows. $(\mathcal{I}-\E)$ is diagonalized by its matrix of orthonormal eigenvectors
\begin{eqnarray}
   \Lambda &\equiv& \mathcal{S}^{\mathsf{T}} (\mathcal{I}-\E) \mathcal{S} \nonumber \\
   &=& \text{diag} \left[0,1,1,1\ldots \right],\\
   \mathcal{S} &=&\left(\vec{v},\vec{u}_1,\vec{u}_2,\ldots \right)
\end{eqnarray}
$\vec{v}$ is defined in \eqref{vmode}, and we will not need the precise form of the $\vec{u}$'s. Under this change of variables, whose jacobian is $1$, $\mathcal{M}$ becomes
\begin{eqnarray} \label{matrix}
	\left(
\begin{array}{cc}
 1 &   \\
   & \mathcal{S}^{\mathsf{T}} \\
\end{array}
\right) \mathcal{M} \left(
\begin{array}{cc}
 1 &    \\
    & \mathcal{S} \\
\end{array}
\right)
	&=& \left(
\begin{array}{cccccc}
 a & \vec{J}\cdot \vec v & \vec{J}\cdot \vec{u}_1 &
   \vec{J}\cdot \vec{u}_2 & \ldots    \\
 \vec{J}\cdot \vec{v} & 0  &    &    &      \\
 \vec{J}\cdot \vec{u}_1 &    & \lambda _1 &    &      \\
 \vec{J}\cdot \vec{u}_2 &    &    & \lambda _2 &      \\
 \vdots  &    &    &    & \ddots   \\
\end{array}
\right)
\end{eqnarray}
We denote our new integration variables $\y_{m'} \equiv (\mathcal{S}^\top)_{m'm} \frac{\delta z_{km}+\delta \bar{z}_{km}}{\sqrt{2}}$, and for the tachyonic mode, $y_0 \equiv \v = \sum_{m=0}^\infty v_m \frac{\delta z_{km}+\delta \bar{z}_{km}}{\sqrt{2}}$. We can now write the quadratic form as follows  
\begin{equation}
	\left(\delta T,\vec{y}^{\mathsf{T}}\right) \mathcal{M} \left(
\begin{array}{c}
 \delta T \\
 \vec{y} \\
\end{array}
\right) =  \vec{y}^{\mathsf{T}} \Lambda  \vec{y} + 2 \delta T \sum _{i=1}^{\infty } y^i \vec{J}\cdot
   \vec{u}_i +  \left(\delta T, v_+ \right) \left(
\begin{array}{cc}
 a & \vec{J}\cdot \vec v \\
 \vec{J}\cdot \vec v & 0 \\
\end{array}
\right) \left(
\begin{array}{c}
 \delta T \\
 v_+ \\
\end{array}
\right)
\end{equation}
The first two terms on the RHS are independent of $y_0$. Thus completing the square in $\y_i$ ($i>0$) to eliminate the second term only has the effect of modifying the coefficient $a$ of $\delta T^2$ appearing in the last term. But due to the form of the latter, the determinant obtained after integrating out $\delta T$, $y_0$ is independent of $a$. 

Therefore the total contribution from integrating out $Re [\delta z_{km}]$ and $\delta T$ is
\begin{eqnarray}
	\text{det}^\prime \left(\mathcal{I}-\E(k)\right)^{-1/2} \det \left[\left(
\begin{array}{cc}
 a  & \vec{J}\cdot \vec v \\
 \vec{J}\cdot \vec v & 0 \\
\end{array}
\right)\right]^{-1/2} &=& \left( {\prod}^\prime 1 \right) \det \left[\left(
\begin{array}{cc}
 a  & \vec{J}\cdot \vec v \\
 \vec{J}\cdot \vec v & 0 \\
\end{array}
\right)\right]^{-1/2} \nonumber\\
	&=& \pm i \left| \vec{J}\cdot \vec v \right|^{-1}\\
    &=& \boxed{ \pm i \sqrt{\frac{4 \pi  \alpha'  T_0^3 \sinh (2 \pi  \epsilon )}{d^2 \pi \epsilon}} }
\end{eqnarray}
where the prime means we exclude the zero mode. The square root of a negative determinant gives rise to a factor $i$. This ``tachyonic'' mode corresponds to fluctuations in the radius of the instanton, with respect to which it is unstable.

\subsection*{Imaginary $n=k$ modes \& zero mode}
Returning to \eqref{tzquadratic}, we note that $Im [\delta z]$ does not couple to $\delta T$, so in this case the zero eigenvalue of $(\mathcal{I}-\E(k))$ corresponds to a genuine zero-mode of the quadratic action:
\begin{equation}
	\z \equiv \sum_{m=0}^\infty v_m \frac{\delta z_{km}-\delta\bar z_{km}}{\sqrt{2}i}
\end{equation} 
The gauge-fixing procedure is described in subsection \ref{zeromode}. The net result is the removal of the zero eigenvalue, and the introduction of a compensating Faddeev-Popov jacobian,
\begin{equation}
	\iint \!\! d\sigma d\tau \left[\dot{\bar{z}}(\sigma, \tau) \dot{\hat z}^{cl}(\sigma,\tau) + \text{c.c.}\right] = \modd{\dot z^{cl}} + \iint \!\! d\sigma d\tau \left[\delta \dot{\bar{z}}\dot{\hat z}^{cl}(\sigma,\tau) + \text{c.c.}\right]
\end{equation}
The first term evaluates to 
$$2\pi k \frac{d}{\sqrt{2} \pi \epsilon} \sqrt{\frac{2\pi\epsilon+\sinh(2\pi\epsilon)}{4\pi\epsilon}},$$
while the latter term, which is the projection of $\delta z$ onto the tachyonic mode $z^{cl}$, is a subleading correction and is to be dropped in the semiclassical approximation.

It is important here to note the following subtlety, that the rescaling of $\delta z_{nm}$ (equation \eqref{gscale}) does not commute with the introduction of our collective coordinate. To account for this, we regulate the determinant, divide out by the (putative) zero eigenvalue, and then take the limit of the regulator going to zero. 

We want the determinant, with zero-eigenvalue excluded, of the operator
$$ \hat L \equiv \frac{1}{2\pi \alpha'T_0} (-\partial_\sigma^2 + (\pi \epsilon)^2)$$
with the appropriate boundary conditions. A possible regularization is 
$$
\hat L^{(\delta)} = \frac{1}{2\pi \alpha'T_0} \left(-\partial_\sigma^2 + (\pi \epsilon (1+\delta))^2 \right)
$$
Then the zero eigenvalue gets shifted to 
$$\lambda_0^{(\delta)}  = \frac{2\pi k\epsilon}{\alpha'}\delta +\mathcal{O}(\delta^2)$$
and $g(n,m)$ (equation \eqref{gscale}) gets modified in accordingly. Now rescale as before, $\delta z_{km} \rightarrow \sqrt{g^{(\delta)}_{km}} \delta z_{km}$. The corresponding jacobian $J_{g}$ has an $\mathcal{O}(\delta)$ correction which will not be important. The resulting quadratic form is now $(\mathcal{I}-\E^{(\delta)}(k))$, where $\E^{(\delta)}$ is defined as in \eqref{Ematrix} but with $g(n,m)$ replaced everywhere by $g^{(\delta)}(n,m)$. This has the same structure as before, with only the pseudo-zero eigenvalue $\tilde{\lambda}_0$ modified to
\begin{align}
	\tilde{\lambda}_0^{(\delta)} = 1-2\pi k E \left[ g^{(\delta)}_{k0} + 2\sum_{m=1}^\infty g^{(\delta)}_{km} \right] &= 1-\frac{\tanh(\pi\epsilon) \coth (\pi \epsilon (1+\delta) ) }{(1+\delta)} \nonumber\\
    &= \left(1+2 \pi  \epsilon\,  \text{csch}(2 \pi  \epsilon )\right) \cdot \delta + \mathcal{O}(\delta^2)
\end{align}
The determinant is then
\begin{eqnarray}
	(\text{det} '\hat{L})^{-1/2} = \lim_{\delta \rightarrow 0}\sqrt{\frac{\lambda_0^{(\delta)}}{\det \hat{L}^{(\delta)}} }
    = \lim_{\delta\rightarrow 0} J_{g}^{(\delta)} \sqrt{\lambda_0^{(\delta)}/\tilde{\lambda}_0^{(\delta)}}
    = J_{g} \left[\frac{2\pi k \epsilon}{\alpha'} \frac{\sinh (2\pi\epsilon)}{2\pi\epsilon + \sinh(2\pi\epsilon)}\right]^{1/2}
\end{eqnarray}
The factor $J_{g}$ was already accounted in \eqref{Jg}, so the net contribution obtained here for the $n=k$ imaginary modes (including the Gribov factor $\omega^{-1}=1/2k$) is
\begin{equation}
	\left[\frac{2\pi k \epsilon}{\alpha'} \frac{\sinh (2\pi\epsilon)}{2\pi\epsilon + \sinh(2\pi\epsilon)}\right]^{1/2} \cdot \frac{1}{2k} \cdot
    \frac{2\pi k d}{\pi \epsilon} \sqrt{\frac{2\pi\epsilon+\sinh(2\pi\epsilon)}{4\pi\epsilon}} = \boxed{\frac{d\, k \sinh(\pi\epsilon)}{\epsilon\sqrt{k\alpha'\tanh(\pi\epsilon)}}}
\end{equation}

\subsection*{Final Result}
Gathering all of the factors, including the product $\prod_{nm} (2\pi) $ evaluated in \eqref{twopiprod}, we obtain for the tunneling amplitude\footnote{
In detail:
\begin{multline}
	\frac{2}{V} \Im (Z) = 2\sum_{k=1}^\infty e^{-\frac{ k d^2}{2 \pi  \alpha'  \varepsilon }} \cdot
    \underbrace{\frac{1}{2T_0}      \vphantom{\left[\frac{T_0}{4\pi^2 \alpha'}\right]^\frac{D-1}{2}}	 }_{\text{measure}}\cdot
	\underbrace{\eta^{-24}(i/2T_0)     \vphantom{\left[\frac{T_0}{4\pi^2 \alpha'}\right]^\frac{D-1}{2}}		}_{\delta x_{a/A}^{m>0}}\cdot
	\underbrace{\left[\frac{T_0}{4\pi^2 \alpha'}\right]^\frac{D-1}{2}}_{\delta x_{a/A}^{m=0}}\cdot 
	\underbrace{\eta^{+2}(i/2T_0)	\vphantom{\left[\frac{T_0}{4\pi^2 \alpha'}\right]^\frac{D-1}{2}}	 }_{\text{ghosts}} \cdot  
	\underbrace{\frac{2\pi}{\eta^2(i/2T_0)}\frac{T_0}{4\pi^2\alpha'}	\vphantom{\left[\frac{T_0}{4\pi^2 \alpha'}\right]^\frac{D-1}{2}}		}_{\delta z ~\text{rescaling}} \cdot \\
	\underbrace{\frac{(-1)^{k+1}e^{-\pi k\varepsilon}}{\cosh^2\pi\varepsilon }}_{\delta z_{n\neq k}}\cdot
	\underbrace{(\pm i) \sqrt{\frac{4 \pi  \alpha'  T_0^3 \sinh (2 \pi  \epsilon )}{d^2 \pi \epsilon}}}_{\text{tachyon}} \cdot
	\underbrace{\frac{d\, k \sinh (\pi  \epsilon )}{\epsilon  \sqrt{k \alpha'  \tanh (\pi  \epsilon )}}}_{J_{FP}} \cdot
\frac{1}{2\pi}
\end{multline}
}
\begin{equation}
	\Gamma_{\text{string}} = \frac{2}{V} \Im (Z) = \pm i \frac{ \tanh\pi\varepsilon}{\varepsilon}
\sum_{k=1}^\infty  \frac{   (-1)^{k+1}  e^{  -\frac{2\pi \alpha' k M^2 }{\varepsilon} -k\pi\varepsilon }      } {
 \eta^{24}(i k/\varepsilon)  }
\left[ \frac{\varepsilon/2\alpha'}{4\pi^2 k } \right]^{ \frac{D+1}{2} }
\end{equation}
in agreement with the Gelfand-Yaglom calculation of section \ref{instanton}.

\section{The Dedekind eta and Jacobi theta functions}  

The Dedekind eta function is defined by 
\begin{equation} \label{dedekindeta}
\eta(\tau) = e^{\pi i\tau/12}\prod_{k=1}^\infty\left(1-e^{2\pi i k\tau}\right)
\end{equation}
It has the property that
\begin{equation} \label{etamod}
	\eta({-1/\tau})=\sqrt{-i\tau}\,\eta(\tau)
\end{equation}
We shall also make use of the Jacobi theta function 
\begin{align}
\Theta_{11}(\nu|\tau)=-2e^{i\frac{\pi\tau}{4}}\sin\pi\nu\prod_{m=1}^\infty(1-e^{2\pi i\tau m})
(1-e^{2\pi i(m\tau+\nu) })(1-e^{2\pi i(m\tau-\nu) })
\end{align}
This theta function has the modular transformations
\begin{align}
\Theta_{11}(\nu|\tau+1)&=e^{i\frac{\pi}{4}}\Theta_{11}(\nu|\tau)  \\
\Theta_{11}(\nu/\tau,-1/\tau)&=-i(-i\tau)^{\frac{1}{2}}e^{\pi i\nu^2/\tau}\Theta_{11}(\nu|\tau)
\end{align}
The definition, further properties and uses of the eta- and theta-functions can be found
in string theory textbooks.  Here, we use the notational conventions of Polchinski \cite{Polchinski2007a}.

\section{Infinite products and \texorpdfstring{$\zeta$}{zeta}-regularization}\label{zeta}

We use zeta-function regularization to define all divergent summations and products in this paper. The Riemann zeta-function is given by
$$
	\zeta(s)~=~\sum_{n=1}^\infty  n^{-s}
$$
This function is defined as a function of the complex variable which is the analytic continuation of 
the function of real $s$ in the domain $s>1$ where it is well-defined.  As a function of complex $s$, it has
a pole on the real axis at $s=1$ and it is finite at negative real values of $s$.  
For completeness we give here explicit computations of some of the infinite products occurring throughout the paper.
The values of the zeta-function of interest to us are
$$
\zeta(0)=-\frac{1}{2}~,~
\zeta(-1)=-\frac{1}{12}~,~
\zeta'(0)=-\frac{1}{2}\ln(2\pi)
$$
Examples of infinite sums and products which we encounter in our computations are
\begin{eqnarray}
	\sum_{m=1}^\infty 1 &\equiv &  \lim_{s\to 0}\zeta(s) = -\frac{1}{2} \\
	\sum_{m=1}^\infty m &\equiv & \lim_{s\to -1}\zeta(s) = -\frac{1}{12} \label{sumofm}\\
    \prod_{n=1}^{\infty} (2\pi n) &\equiv & \lim_{s\to 0}(2\pi)^{\zeta(s)} \cdot\lim_{\bar s\to 0} e^{-\frac{d}{d\bar s}\zeta(\bar s) }     = 1 \\
    \nplus \Omega_n &=& \nplus (2\pi n) \frac{ \varepsilon}{2k} = \sqrt{\frac{2k}{\varepsilon}} 
\end{eqnarray}
Using the above regularizations, and the product formula
\begin{align}\label{productformula}
\prod_{n=1}^\infty \left[1+\frac{a^2}{n^2}\right]=\frac{ \sinh\pi a}{ \pi a}
\end{align}
we find
\begin{eqnarray*}
	\prod_{n=-\infty}^\infty \left[ (\pi n)^2 a^2 + (\pi m)^2 b^2 \right] &=& (\pi m b)^2 \prod_{n>0} \left[ (\pi n)^2 a^2 + (\pi m)^2 b^2 \right]^2 \\
    &=& (\pi m b)^2 \prod_{n>0} (\pi n a)^4 \prod_{n>0} \left[ 1+ \left(\frac{\pi m b}{\pi n a} \right)^2 \right]^2 \\
    &=& (\pi m b)^2 \left(\frac{2}{a}\right)^2 \frac{\sinh^2(\pi m b/a)}{(\pi m b /a)^2} \\
    &=& 4 \sinh^2 (\pi m b/a)
\end{eqnarray*}
By \eqref{dedekindeta} and \eqref{sumofm} we further have that
\begin{eqnarray*}
	\prod_{m\geq 1} 2\sinh (\pi m b/a) &=& e^{\pi b/a\sum_{m=1}^\infty m} \prod_{m\geq 1} \left( 1-e^{-2\pi mba}\right)\\
    &=& e^{-\frac{1}{12}\pi b/a} \prod_{m=1}^\infty \left(1-e^{-2\pi m b/a}\right) \\
    &=& \eta(ib/a) = \sqrt{\frac{a}{b}}\eta(ia/b).
\end{eqnarray*}
In section (\ref{GYevaluation}) and appendix (\ref{neumannmodeexpansion}) one encounters the following infinite product:
\begin{eqnarray} \label{appendixmodproduct}
	\prod_{n\neq 0,k}\frac{1}{\det\left({\cal I}-\E(n)\right)}&=&\frac{1}{\prod_{n\neq 0,k} \left[ 1-\tanh\pi\varepsilon\coth \pi \varepsilon n/k\right] } \nonumber \\
	&=&\lim_{\varkappa\to k} \frac{\left[ 1-\tanh\pi\varepsilon\coth \pi \varepsilon k/\varkappa\right] }
{\prod_{n=1}^\infty\left[ 1-\tanh^2\pi\varepsilon\coth^2 \pi \varepsilon n/\varkappa\right]  }\nonumber \\
	&=&\lim_{\varkappa\to k} 
{\left[ 1-\tanh\pi\varepsilon\coth \pi \varepsilon k/\varkappa \right] }
\,\frac{\varkappa}{\varepsilon}\,\frac{\tanh\pi\varepsilon \;e^{-\pi\varkappa \varepsilon}}{\sin(\pi\varkappa)} \cdot \nonumber\\
&& \phantom{\lim_{\varkappa\to k} }\cdot \prod_{n=1}^\infty  \frac{[1-e^{-2\pi \varkappa n/\varepsilon}]^2}{
(1-e^{-2\pi\varkappa(n+i\varepsilon)/\varepsilon })(1-e^{-2\pi\varkappa (n-i\varepsilon)/\epsilon })  }\nonumber\\
&=&\boxed{\frac{(-1)^{k+1}e^{-\pi k\varepsilon}}{\cosh^2\pi\varepsilon }}
\end{eqnarray}
In the third line we performed a modular transformation as in equation (\ref{modularproduct}), and for the last step observed that the remaining infinite product is regular for all $\varkappa>0$.

\bibliographystyle{utphys}
\bibliography{SchwingerRefs}

\providecommand{\href}[2]{#2}\begingroup\raggedright\begin{thebibliography}{10}

\bibitem{Schwinger1951}
J.~S. Schwinger, ``{On gauge invariance and vacuum polarization},''
\href{http://dx.doi.org/10.1103/PhysRev.82.664}{{\em Phys. Rev.} {\bfseries 82}
  (1951) 664--679}.

\bibitem{Bachas1992}
C.~Bachas and M.~Porrati, ``{Pair creation of open strings in an electric
  field},'' \href{http://dx.doi.org/10.1016/0370-2693(92)90806-F}{{\em Phys.
  Lett.} {\bfseries B296} (1992) 77--84},
\href{http://arxiv.org/abs/hep-th/9209032}{{\ttfamily arXiv:hep-th/9209032
  [hep-th]}}.

\bibitem{Ambjorn2003}
J.~Ambjorn, Y.~M. Makeenko, G.~W. Semenoff, and R.~J. Szabo, ``{String theory
  in electromagnetic fields},''
  \href{http://dx.doi.org/10.1088/1126-6708/2003/02/026}{{\em JHEP} {\bfseries
  02} (2003) 026},
\href{http://arxiv.org/abs/hep-th/0012092}{{\ttfamily arXiv:hep-th/0012092
  [hep-th]}}.

\bibitem{Fradkin1985}
E.~S. Fradkin and A.~A. Tseytlin, ``{Nonlinear Electrodynamics from Quantized
  Strings},''
\href{http://dx.doi.org/10.1016/0370-2693(85)90205-9}{{\em Phys. Lett.}
  {\bfseries 163B} (1985) 123--130}.

\bibitem{Fradkin1985a}
E.~S. Fradkin and A.~A. Tseytlin, ``{Effective Field Theory from Quantized
  Strings},''
\href{http://dx.doi.org/10.1016/0370-2693(85)91190-6}{{\em Phys. Lett.}
  {\bfseries 158B} (1985) 316--322}.

\bibitem{Fradkin1985b}
E.~S. Fradkin and A.~A. Tseytlin, ``{Quantum String Theory Effective Action},''
  \href{http://dx.doi.org/10.1016/0550-3213(86)90522-5,
  10.1016/0550-3213(85)90559-0}{{\em Nucl. Phys.} {\bfseries B261} (1985)
  1--27}.
[Erratum: Nucl. Phys.B269,745(1986)].

\bibitem{Burgess1987}
C.~P. Burgess, ``{Open String Instability in Background Electric Fields},''
\href{http://dx.doi.org/10.1016/0550-3213(87)90590-6}{{\em Nucl. Phys.}
  {\bfseries B294} (1987) 427--444}.

\bibitem{Billo1997}
M.~Billo, P.~Di~Vecchia, and D.~Cangemi, ``{Boundary states for moving
  D-branes},'' \href{http://dx.doi.org/10.1016/S0370-2693(97)00329-8}{{\em
  Phys. Lett.} {\bfseries B400} (1997) 63--70},
\href{http://arxiv.org/abs/hep-th/9701190}{{\ttfamily arXiv:hep-th/9701190
  [hep-th]}}.

\bibitem{Gordon2015}
J.~Gordon and G.~W. Semenoff, ``{World-line instantons and the Schwinger effect
  as a Wentzel-Kramers-Brillouin exact path integral},''
  \href{http://dx.doi.org/10.1063/1.4908556}{{\em J. Math. Phys.} {\bfseries
  56} (2015) 022111},
\href{http://arxiv.org/abs/1407.0987}{{\ttfamily arXiv:1407.0987 [hep-th]}}.

\bibitem{Gordon2016}
J.~Gordon and G.~W. Semenoff, ``{Schwinger pair production: Explicit
  Localization of the world-line instanton},''
\href{http://arxiv.org/abs/1612.05909}{{\ttfamily arXiv:1612.05909 [hep-th]}}.

\bibitem{Schubert2010}
C.~Schubert and A.~Torrielli, ``{Open string pair creation from worldsheet
  instantons},'' \href{http://dx.doi.org/10.1088/1751-8113/43/40/402003}{{\em
  J. Phys.} {\bfseries A43} (2010) 402003},
\href{http://arxiv.org/abs/1008.2068}{{\ttfamily arXiv:1008.2068 [hep-th]}}.

\bibitem{Ilderton2014}
A.~Ilderton, ``{Localisation in worldline pair production and lightfront
  zero-modes},'' \href{http://dx.doi.org/10.1007/JHEP09(2014)166}{{\em JHEP}
  {\bfseries 09} (2014) 166},
\href{http://arxiv.org/abs/1406.1513}{{\ttfamily arXiv:1406.1513 [hep-th]}}.

\bibitem{Abouelsaood1987}
A.~Abouelsaood, C.~G. Callan, Jr., C.~R. Nappi, and S.~A. Yost, ``{Open Strings
  in Background Gauge Fields},''
\href{http://dx.doi.org/10.1016/0550-3213(87)90164-7}{{\em Nucl. Phys.}
  {\bfseries B280} (1987) 599--624}.

\bibitem{Polchinski2007}
J.~Polchinski, {\em {String theory. Vol. 2: Superstring theory and beyond}}.
\newblock Cambridge University Press,
2007.
\newblock

\bibitem{Gelfand1960}
I.~M. Gelfand and A.~M. Yaglom, ``{Integration in functional spaces and it
  applications in quantum physics},''
\href{http://dx.doi.org/10.1063/1.1703636}{{\em J. Math. Phys.} {\bfseries 1}
  (1960) 48}.

\bibitem{Forman1987}
R.~Forman, ``Functional determinants and geometry,''
  \href{http://dx.doi.org/10.1007/BF01391828}{{\em Inventiones mathematicae}
  {\bfseries 88} no.~3, (Oct., 1987) 447}.
  \url{http://dx.doi.org/10.1007/BF01391828}.

\bibitem{McKane1995}
A.~J. McKane and M.~B. Tarlie, ``{Regularization of functional determinants
  using boundary perturbations},''
  \href{http://dx.doi.org/10.1088/0305-4470/28/23/032}{{\em J. Phys.}
  {\bfseries A28} (1995) 6931--6942},
\href{http://arxiv.org/abs/cond-mat/9509126}{{\ttfamily arXiv:cond-mat/9509126
  [cond-mat]}}.

\bibitem{Kirsten2003}
K.~Kirsten and A.~J. McKane, ``{Functional determinants by contour integration
  methods},'' \href{http://dx.doi.org/10.1016/S0003-4916(03)00149-0}{{\em
  Annals Phys.} {\bfseries 308} (2003) 502--527},
\href{http://arxiv.org/abs/math-ph/0305010}{{\ttfamily arXiv:math-ph/0305010
  [math-ph]}}.

\bibitem{Kirsten2004}
K.~Kirsten and A.~J. McKane, ``{Functional determinants for general
  Sturm-Liouville problems},''
  \href{http://dx.doi.org/10.1088/0305-4470/37/16/014}{{\em J. Phys.}
  {\bfseries A37} (2004) 4649--4670},
\href{http://arxiv.org/abs/math-ph/0403050}{{\ttfamily arXiv:math-ph/0403050
  [math-ph]}}.

\bibitem{Kirsten2008}
K.~Kirsten and P.~Loya, ``{Computation of determinants using contour
  integrals},'' \href{http://dx.doi.org/10.1119/1.2794348}{{\em Am. J. Phys.}
  {\bfseries 76} (2008) 60--64},
\href{http://arxiv.org/abs/0707.3755}{{\ttfamily arXiv:0707.3755 [hep-th]}}.

\bibitem{Kirsten2010}
K.~Kirsten, ``{Functional determinants in higher dimensions using contour
  integrals},''
\href{http://arxiv.org/abs/1005.2595}{{\ttfamily arXiv:1005.2595 [hep-th]}}.

\bibitem{Dunne2008}
G.~V. Dunne, ``{Functional determinants in quantum field theory},''
  \href{http://dx.doi.org/10.1088/1751-8113/41/30/304006}{{\em J. Phys.}
  {\bfseries A41} (2008) 304006},
\href{http://arxiv.org/abs/0711.1178}{{\ttfamily arXiv:0711.1178 [hep-th]}}.

\bibitem{Durin2003}
B.~Durin and B.~Pioline, ``{Open strings in relativistic ion traps},''
  \href{http://dx.doi.org/10.1088/1126-6708/2003/05/035}{{\em JHEP} {\bfseries
  05} (2003) 035},
\href{http://arxiv.org/abs/hep-th/0302159}{{\ttfamily arXiv:hep-th/0302159
  [hep-th]}}.

\bibitem{Bolognesi2016}
S.~Bolognesi, E.~Rabinovici, and G.~Tallarita, ``{String pair production in non
  homogeneous backgrounds},''
  \href{http://dx.doi.org/10.1007/JHEP04(2016)174}{{\em JHEP} {\bfseries 04}
  (2016) 174},
\href{http://arxiv.org/abs/1601.04758}{{\ttfamily arXiv:1601.04758 [hep-th]}}.

\bibitem{Condeescu2017}
C.~Condeescu, E.~Dudas, and G.~Pradisi, ``{Open Strings and Electric Fields in
  Compact Spaces},''
\href{http://arxiv.org/abs/1705.02352}{{\ttfamily arXiv:1705.02352 [hep-th]}}.

\bibitem{Bachas1996}
C.~Bachas, ``{D-brane dynamics},''
  \href{http://dx.doi.org/10.1016/0370-2693(96)00238-9}{{\em Phys. Lett.}
  {\bfseries B374} (1996) 37--42},
\href{http://arxiv.org/abs/hep-th/9511043}{{\ttfamily arXiv:hep-th/9511043
  [hep-th]}}.

\bibitem{Bachlechner2013}
T.~C. Bachlechner and L.~McAllister, ``{D-brane Bremsstrahlung},''
  \href{http://dx.doi.org/10.1007/JHEP10(2013)022}{{\em JHEP} {\bfseries 10}
  (2013) 022},
\href{http://arxiv.org/abs/1306.0003}{{\ttfamily arXiv:1306.0003 [hep-th]}}.

\bibitem{DAmico2015}
G.~D'Amico, R.~Gobbetti, M.~Kleban, and M.~Schillo, ``{D-brane scattering and
  annihilation},'' \href{http://dx.doi.org/10.1007/JHEP01(2015)050}{{\em JHEP}
  {\bfseries 01} (2015) 050},
\href{http://arxiv.org/abs/1408.2540}{{\ttfamily arXiv:1408.2540 [hep-th]}}.

\bibitem{Semenoff2011}
G.~W. Semenoff and K.~Zarembo, ``{Holographic Schwinger Effect},''
  \href{http://dx.doi.org/10.1103/PhysRevLett.107.171601}{{\em Phys. Rev.
  Lett.} {\bfseries 107} (2011) 171601},
\href{http://arxiv.org/abs/1109.2920}{{\ttfamily arXiv:1109.2920 [hep-th]}}.

\bibitem{Faulkner2009}
T.~Faulkner and H.~Liu, ``{Meson widths from string worldsheet instantons},''
  \href{http://dx.doi.org/10.1016/j.physletb.2009.01.071}{{\em Phys. Lett.}
  {\bfseries B673} (2009) 161--165},
\href{http://arxiv.org/abs/0807.0063}{{\ttfamily arXiv:0807.0063 [hep-th]}}.

\bibitem{Basar2012a}
G.~Basar, D.~E. Kharzeev, H.-U. Yee, and I.~Zahed, ``{Holographic Pomeron and
  the Schwinger Mechanism},''
  \href{http://dx.doi.org/10.1103/PhysRevD.85.105005}{{\em Phys. Rev.}
  {\bfseries D85} (2012) 105005},
\href{http://arxiv.org/abs/1202.0831}{{\ttfamily arXiv:1202.0831 [hep-th]}}.

\bibitem{Tseytlin1999}
A.~A. Tseytlin, ``{Born-Infeld action, supersymmetry and string theory},''
\href{http://arxiv.org/abs/hep-th/9908105}{{\ttfamily arXiv:hep-th/9908105
  [hep-th]}}.

\bibitem{Callan1990}
C.~G. Callan, Jr. and L.~Thorlacius, ``{Open String Theory as Dissipative
  Quantum Mechanics},''
\href{http://dx.doi.org/10.1016/0550-3213(90)90060-Q}{{\em Nucl. Phys.}
  {\bfseries B329} (1990) 117--138}.

\bibitem{Caldeira1983}
A.~O. Caldeira and A.~J. Leggett, ``{Quantum tunneling in a dissipative
  system},''
\href{http://dx.doi.org/10.1016/0003-4916(83)90202-6}{{\em Annals Phys.}
  {\bfseries 149} (1983) 374--456}.

\bibitem{Caldeira1983a}
A.~O. Caldeira and A.~J. Leggett, ``{Path integral approach to quantum Brownian
  motion},''
\href{http://dx.doi.org/10.1016/0378-4371(83)90013-4}{{\em Physica} {\bfseries
  121A} (1983) 587--616}.

\bibitem{Acatrinei1999}
C.~Acatrinei and R.~Iengo, ``{Pair production of open strings: Relativistic
  versus dissipative dynamics},''
  \href{http://dx.doi.org/10.1016/S0550-3213(98)00734-2}{{\em Nucl. Phys.}
  {\bfseries B539} (1999) 513--532},
\href{http://arxiv.org/abs/hep-th/9806048}{{\ttfamily arXiv:hep-th/9806048
  [hep-th]}}.

\bibitem{Polchinski2007a}
J.~Polchinski, {\em {String theory. Vol. 1: An introduction to the bosonic
  string}}.
\newblock Cambridge University Press,
2007.
\newblock

\end{thebibliography}\endgroup

\end{document}